

\input harvmac
\newif\ifdraft

\noblackbox
\catcode`\@=11
\newif\iffrontpage
\def\figin{\epsfcheck\figin}\def\figins{\epsfcheck\figins}
\def\epsfcheck{\ifx\epsfbox\UnDeFiNeD
\message{(NO epsf.tex, FIGURES WILL BE IGNORED)}
\gdef\figin##1{\vskip2in}\gdef\figins##1{\hskip.5in}%
\else\message{(FIGURES WILL BE INCLUDED)}%
\gdef\figin##1{##1}\gdef\figins##1{##1}\fi}
\def\DefWarn#1{}
\def\figinsert{\goodbreak\midinsert}
\def\ifig#1#2#3{\DefWarn#1\xdef#1{fig.~\the\figno}
\writedef{#1\leftbracket fig.\noexpand~\the\figno}%
\figinsert\figin{\centerline{#3}}\medskip%
\centerline{\vbox{\baselineskip12pt
\advance\hsize by -1truein\noindent\footnotefont%
\centerline{{\bf Fig.~\the\figno}~#2}}
}\bigskip\endinsert\global\advance\figno by1}
\ifx\answ\bigans
\def\titleft{\titsm}
\magnification=1200\baselineskip=14pt plus 2pt minus 1pt
%
\advance\hoffset by-0.075truein
\hsize=6.15truein\vsize=600.truept\hsbody=\hsize\hstitle=\hsize
\else\let\lr=L
\def\titleft{\titla}
\magnification=1000\baselineskip=14pt plus 2pt minus 1pt
%
\hoffset=-.48truein\voffset=-.1truein
\vsize=6.5truein
\hstitle=8.truein\hsbody=4.75truein
\fullhsize=10truein\hsize=\hsbody
\fi
%
\parskip=4pt plus 15pt minus 1pt

\font\titla=cmr10 scaled\magstep3
\font\tenmss=cmss10
\font\absmss=cmss10 scaled\magstep1

\font\twelvebf=cmbx10 scaled\magstep1

\newfam\mssfam
\font\footrm=cmr8  \font\footrms=cmr5
\font\footrmss=cmr5   \font\footi=cmmi8
\font\footis=cmmi5   \font\footiss=cmmi5
\font\footsy=cmsy8   \font\footsys=cmsy5
\font\footsyss=cmsy5   \font\footbf=cmbx8
\font\footmss=cmss8
\def\footfont{\def\rm{\fam0\footrm}
\textfont0=\footrm \scriptfont0=\footrms
\scriptscriptfont0=\footrmss
\textfont1=\footi \scriptfont1=\footis
\scriptscriptfont1=\footiss
\textfont2=\footsy \scriptfont2=\footsys
\scriptscriptfont2=\footsyss
\textfont\itfam=\footi \def\it{\fam\itfam\footi}
\textfont\mssfam=\footmss \def\mss{\fam\mssfam\footmss}
\textfont\bffam=\footbf \def\bf{\fam\bffam\footbf} \rm}
\def\tenpoint{\def\rm{\fam0\tenrm}
\textfont0=\tenrm \scriptfont0=\sevenrm
\scriptscriptfont0=\fiverm
\textfont1=\teni  \scriptfont1=\seveni
\scriptscriptfont1=\fivei
\textfont2=\tensy \scriptfont2=\sevensy
\scriptscriptfont2=\fivesy
\textfont\itfam=\tenit \def\it{\fam\itfam\tenit}
\textfont\mssfam=\tenmss \def\mss{\fam\mssfam\tenmss}
\textfont\bffam=\tenbf \def\bf{\fam\bffam\tenbf} \rm}
\ifx\answ\bigans\def\abstractfont{\tenpoint}\else
\def\abstractfont{\def\rm{\fam0\absrm}
\textfont0=\absrm \scriptfont0=\absrms
\scriptscriptfont0=\absrmss
\textfont1=\absi \scriptfont1=\absis
\scriptscriptfont1=\absiss
\textfont2=\abssy \scriptfont2=\abssys
\scriptscriptfont2=\abssyss
\textfont\itfam=\bigit \def\it{\fam\itfam\bigit}
\textfont\mssfam=\absmss \def\mss{\fam\mssfam\absmss}
\textfont\bffam=\absbf \def\bf{\fam\bffam\absbf}\rm}\fi
%
\def\f@@t{\baselineskip10pt\lineskip0pt\lineskiplimit0pt
\bgroup\aftergroup\@foot\let\next}
\setbox\strutbox=\hbox{\vrule height 8.pt depth 3.5pt width\z@}
\def\vfootnote#1{\insert\footins\bgroup
\baselineskip10pt\footfont
\interlinepenalty=\interfootnotelinepenalty
\floatingpenalty=20000
\splittopskip=\ht\strutbox \boxmaxdepth=\dp\strutbox
\leftskip=24pt \rightskip=\z@skip
\parindent=12pt \parfillskip=0pt plus 1fil
\spaceskip=\z@skip \xspaceskip=\z@skip
\Textindent{$#1$}\footstrut\futurelet\next\fo@t}
\def\Textindent#1{\noindent\llap{#1\enspace}\ignorespaces}
\def\footnote#1{\attach{#1}\vfootnote{#1}}%

\def\foot{\attach\footsymbolgen\vfootnote{\footsymbol}}
\let\footsymbol=\star
\newcount\lastf@@t           \lastf@@t=-1
\newcount\footsymbolcount    \footsymbolcount=0
\def\footsymbolgen{\relax\footsym
\global\lastf@@t=\pageno\footsymbol}
\def\footsym{\ifnum\footsymbolcount<0
\global\footsymbolcount=0\fi
{\iffrontpage \else \advance\lastf@@t by 1 \fi
\ifnum\lastf@@t<\pageno \global\footsymbolcount=0
\else \global\advance\footsymbolcount by 1 \fi }
\ifcase\footsymbolcount \fd@f\star\or
\fd@f\dagger\or \fd@f\ast\or
\fd@f\ddagger\or \fd@f\natural\or
\fd@f\diamond\or \fd@f\bullet\or
\fd@f\nabla\else \fd@f\dagger
\global\footsymbolcount=0 \fi }
\def\fd@f#1{\xdef\footsymbol{#1}}
\def\space@ver#1{\let\@sf=\empty \ifmmode #1\else \ifhmode
\edef\@sf{\spacefactor=\the\spacefactor}
\unskip${}#1$\relax\fi\fi}
\def\attach#1{\space@ver{\strut^{\mkern 2mu #1}}\@sf}
%
\newif\ifnref
\def\rrr#1#2{\relax\ifnref\nref#1{#2}\else\ref#1{#2}\fi}
\def\ldf#1#2{\begingroup\obeylines
\gdef#1{\rrr{#1}{#2}}\endgroup\unskip}
\def\nrf#1{\nreftrue{#1}\nreffalse}
\def\doubref#1#2{\refs{{#1},{#2}}}

\nreffalse
\def\refout{\listrefs}
%
\def\eqn#1{\xdef #1{(\secsym\the\meqno)}
\writedef{#1\leftbracket#1}%
\global\advance\meqno by1\eqno#1\eqlabeL#1}
\def\eqnalign#1{\xdef #1{(\secsym\the\meqno)}
\writedef{#1\leftbracket#1}%
\global\advance\meqno by1#1\eqlabeL{#1}}
%
\def\chap#1{\global\advance\secno by1\message{(\the\secno\ #1)}
\global\subsecno=0\eqnres@t\noindent{\twelvebf\the\secno\ #1}
\writetoca{{\secsym} {#1}}\par\nobreak\medskip\nobreak}
\def\eqnres@t{\xdef\secsym%
{\the\secno.}\global\meqno=1\bigbreak\bigskip}
\def\sequentialequations{\def\eqnres@t{\bigbreak}}\xdef\secsym{}
\global\newcount\subsecno \global\subsecno=0
\def\sect#1{\global\advance\subsecno
by1\message{(\secsym\the\subsecno. #1)}
\ifnum\lastpenalty>9000\else\bigbreak\fi
\noindent{\bf\secsym\the\subsecno\ #1}\writetoca{\string\quad
{\secsym\the\subsecno.} {#1}}\par\nobreak\medskip\nobreak}
\def\chapter#1{\chap{#1}}
\def\section#1{\sect{#1}}
\def\\{\ifnum\lastpenalty=-10000\relax
\else\hfil\penalty-10000\fi\ignorespaces}
\def\note#1{\leavevmode%
\edef\@@marginsf{\spacefactor=\the\spacefactor\relax}%
\ifdraft\strut\vadjust{%
\hbox to0pt{\hskip\hsize%
\ifx\answ\bigans\hskip.1in\else\hskip .1in\fi%
\vbox to0pt{\vskip-\dp
\strutbox\sevenbf\baselineskip=8pt plus 1pt minus 1pt%
\ifx\answ\bigans\hsize=.7in\else\hsize=.35in\fi%
\tolerance=5000 \hbadness=5000%
\leftskip=0pt \rightskip=0pt \everypar={}%
\raggedright\parskip=0pt \parindent=0pt%
\vskip-\ht\strutbox\noindent\strut#1\par%
\vss}\hss}}\fi\@@marginsf\kern-.01cm}
\def\titlepage{%
\frontpagetrue\nopagenumbers\abstractfont%
\hsize=\hstitle\rightline{\vbox{\baselineskip=10pt%
{\abstractfont\pubnum}}}\pageno=0}
\frontpagefalse
\def\pubnum{}
\def\pdate{\number\month/\number\yearltd}
\def\makefootline{\iffrontpage\vskip .27truein
\line{\the\footline}
\vskip -.1truein\leftline{\vbox{\baselineskip=10pt%
{\abstractfont\pdate}}}
\else\vskip.5cm\line{\hss \tenrm $-$ \folio\ $-$ \hss}\fi}
\def\title#1{\vskip .7truecm\titlestyle{\titleft #1}}
\def\titlestyle#1{\par\begingroup \interlinepenalty=9999
\leftskip=0.02\hsize plus 0.23\hsize minus 0.02\hsize
\rightskip=\leftskip \parfillskip=0pt
\hyphenpenalty=9000 \exhyphenpenalty=9000
\tolerance=9999 \pretolerance=9000
\spaceskip=0.333em \xspaceskip=0.5em
\noindent #1\par\endgroup }
\def\autskip{\ifx\answ\bigans\vskip.5truecm\else\vskip.1cm\fi}
\def\author#1{\vskip .7in \centerline{#1}}

\def\address#1{\ifx\answ\bigans\vskip.2truecm
\else\vskip.1cm\fi{\it \centerline{#1}}}
\def\abstract#1{
\vskip .5in\vfil\centerline
{\bf Abstract}\penalty1000
{{\smallskip\ifx\answ\bigans\leftskip 2pc \rightskip 2pc
\else\leftskip 5pc \rightskip 5pc\fi
\noindent\abstractfont \baselineskip=12pt
{#1} \smallskip}}
\penalty-1000}
\def\endpage{\tenpoint\supereject\global\hsize=\hsbody%
\frontpagefalse\footline={\hss\tenrm\folio\hss}}
%

\def\a{{\alpha}} \def\b{{\beta}} \def\d{{\delta}}
\def\e{{\epsilon}} \def\c{{\gamma}}
\def\G{{\Gamma}}  \def\l{{\lambda}}
\def\L{{\Lambda}} \def\s{{\sigma}} \def\S{{\Sigma}}
\def\cA{{\cal A}} 
\def\cC{{\cal C}} \def\cD{{\cal D}}
\def\cF{{\cal F}}

\def\cL{{\cal L}} \def\cM{{\cal M}}
\def\cN{{\cal N}} 
 \def\cQ{{\cal Q}} \def\cU{{\cal U}}
 \def\cV{{\cal V}}
 \def\cS{{\cal S}} \def\cT{{\cal T}}
\def\IGa{\ralax{{\rm I}\kern-.18em \Gamma}}
\def\IZ{{\hbox{{\rm Z}\kern-.4em\hbox{\rm Z}}}}
\def\IR{{\hbox{{\rm I}\kern-.4em\hbox{\rm R}}}}
\def\IC{{\hbox{{\rm l}\kern-.4em\hbox{\rm C}}}}
\def\bigone{{\hbox{1\kern -.23em{\rm l}}}}
\def\Im{{\rm Im ~}}
\def\Re{{\rm Re ~}}
\def\nup#1({Nucl.\ Phys.\ $\us {B#1}$\ (}
\def\plt#1({Phys.\ Lett.\ $\us  {#1}$\ (}
\def\cmp#1({Comm.\ Math.\ Phys.\ $\us  {#1}$\ (}
\def\prp#1({Phys.\ Rep.\ $\us  {#1}$\ (}
\def\prl#1({Phys.\ Rev.\ Lett.\ $\us  {#1}$\ (}
\def\prv#1({Phys.\ Rev.\ $\us  {#1}$\ (}
\def\mpl#1({Mod.\ Phys.\ Let.\ $\us  {A#1}$\ (}
\def\ijmp#1({Int.\ J.\ Mod.\ Phys.\ $\us {A#1}$\ (}
\def\cqg#1({Class.\ Quantum Grav.\ $\us {#1}$\ (}
\def\anp#1({Ann.\ of Phys.\ $\us {#1}$\ (}
\def\tmp#1({Theor.\ Math.\ Phys.\ $\us {#1}$\ (}

\def\tit#1|{{\it #1},\ }
\def\ft#1#2{{\textstyle{#1\over#2}}}
%
\def\ni{\noindent}
\def\tilde{\widetilde}
\def\bar{\overline}
\def\us#1{\underline{#1}}

\def\hat{\widehat}
\def\to{\rightarrow}
\def\notin{\hbox{{$\in$}\kern-.51em\hbox{/}}}

\def\del{\partial}

 \def\ie{{\it i.e.}\ }
\catcode`\@=12

\def\cy{Calabi--Yau\ }
\def\K{K\"ahler\ }
\def\bF{{\bar F}}

\def\bX{{\bar X}}
\def\ib{{\bar \imath}}
\def\jb{{\bar \jmath}}

\def\aff#1#2{\centerline{$^{#1}${\it #2}}}
\hfill{CERN-TH 7547/94 }\vskip -.2truecm

\hfill{POLFIS-TH. 01/95} \vskip -.2truecm

\hfill{UCLA 94/TEP/45}  \vskip -.2truecm

\hfill{KUL-TF-95/4} \vskip -.2truecm

\hfill{hep-th/9502072 }   \vskip -1truecm

\def\pdate{February 1995}
\titlepage
\vskip .2truecm
\title
 {Duality Transformations in Supersymmetric
Yang--Mills Theories coupled to Supergravity }
\foot{Supported in part by DOE
 grants DE-AC0381-ER50050 and DOE-AT03-88ER40384,Task E. and by EEC
 Science Program SC1*CI92-0789.}
\vskip-.8cm
\author{
A.\ Ceresole$^{1}$, R.\ D'Auria$^{1}$,
S.\ Ferrara$^{2}$ and A.\ Van Proeyen$^{3}$\foot{Onderzoeksleider,
NFWO, Belgium}}
\vskip0.4truecm
\aff1{Dipartimento di Fisica, Politecnico di Torino,}
\centerline{\it  Corso Duca Degli Abruzzi 24, 10129 Torino, Italy}
\centerline{\it and}
\centerline{\it INFN, Sezione di Torino, Italy}
\vskip0.3truecm
\aff2{CERN, 1211 Geneva 23, Switzerland}
\vskip0.3truecm
\aff3{Instituut voor Theoretische Fysica, K.U. Leuven}
\centerline{\it Celestijnenlaan 200 D}
\centerline{\it B--3001 Leuven, Belgium}
\vskip-.8 cm
\def\abs
{\ni
We consider duality transformations in $N=2$, $d=4$ Yang--Mills theory
coupled to $N=2$ supergravity. A symplectic and coordinate covariant
framework is established, which allows one to discuss stringy
`classical and quantum duality symmetries' (monodromies),
incorporating $T$ and $S$ dualities. In particular, we shall be able to
study theories (like $N=2$ heterotic strings) which are
formulated in symplectic basis where a `holomorphic prepotential'
$F$ does not exist, and yet give general expressions for all
relevant physical quantities.  Duality transformations
and symmetries for the $N=1$ matter coupled Yang--Mills supergravity
system are also exhibited. The implications of duality symmetry on all
$N>2$ extended supergravities are briefly mentioned.
We finally give the general form of the central charge and the $N=2$
semiclassical spectrum of the dyonic BPS saturated states
(as it comes by truncation of the $N=4$ spectrum).
}
\abstract{\abs}
\vfill
\endpage
\baselineskip=14pt plus 2pt minus 1pt
\ldf\seiwit{N.\ Seiberg and E.\ Witten, \nup426 (1994) 19,
hep-th/9407087; \nup431 (1994) 484, hep-th/9408099.}
\ldf\KLTY{A.\ Klemm, W.\ Lerche, S.\ Yankielowicz and S. Theisen,
\tit Simple Singularities
and $N=2$ Supersymmetric Yang-Mills Theory | preprint
CERN-TH.7495/94,
 LMU-TPW 94/16, hep-th/9411048 and \tit
On the monodromies of N=2 supersymmetric Yang-Mills theory |  to be
published in the proceedings of the
Workshop on Physics from the Planck Scale to
Electromagnetic Scale, Warsaw, Poland, Sep 21-24, 1994 ,
and for 28th
International Symposium on Particle Theory, Wendisch-Rietz, Germany,
30 Aug - 3 Sep 1994,
preprint CERN-TH-7538-94,
 hep-th/9412158;
P.\ C.\ Argyres and A.\ E.\ Faraggi,
\tit The Vacuum Structure and Spectrum of $N=2$ Supersymmetric
$SU(n)$
Gauge Theory | preprint IASSNS-HEP-94/94, hep-th/9411057}
\ldf\NOI{A.\ Ceresole, R.\ D'Auria and S.\ Ferrara, \plt339B (1994)
71,
hep-th/9408036 .}
\ldf\salsez{ see for instance
A.\ Salam and E.\ Sezgin, \tit Supergravities in diverse
dimensions | World Scientific (1989); L. Castellani, R. D'Auria
and P.\ Fr\`e, \tit Supergravity and Superstrings, a geometric perspective |
World Scientific (1991). }
\ldf\DFF{R.\ D'Auria, S.\ Ferrara and P.\ Fr\`e, \nup359 (1991) 705.}
\ldf\FRSO{P.\ Fr\`e and P.\ Soriani, \nup371 (1992) 659.}
\ldf\dive{P.\ Di Vecchia, R.\  Musto, F.\ Nicodemi and R.\ Pettorino,
\nup252 (1985) 635.}
\ldf\AMAT{D.\  Amati, K.\ Konishi, Y.\ Meurice, G.\ C.\  Rossi
 and G.\  Veneziano, \prp162 (1988) 169.}
\ldf\FSZ{S.\ Ferrara, J.\ Scherk and B.\ Zumino, \nup121 (1977) 393.}
\ldf\jhs{J.\ H.\ Schwarz,
 \tit Evidence for Non--perturbative String Symmetries |  proceedings
 of the
Conference on Topology, Strings and Integrable Models (Satellite
Colloquium to the ICMP-11 18-23 Jul 1994), Paris, France, 25-28 Jul
1994, preprint CALT-68-1965, hep-th/9411178. }
\ldf\GAZU{M.\ K.\ Gaillard and B.\ Zumino, \nup193 (1981) 221.}
\ldf\WVP{B.\ de Wit, F.\ Vanderseypen and A.\ Van Proeyen, \nup400
(1993) 463.}
\ldf\DKL{L.\ J.\ Dixon, V.\ S.\  Kaplunovsky and J. Louis, \nup355
(1990) 27.}
\ldf\CarOvr{G.\ L.\ Cardoso and B.\ A.\ Ovrut, \nup369 (1992) 351.}
\ldf\AGNT{I.\ Antoniadis, E.\ Gava,
K.\ S.\  Narain and T.\ R.\  Taylor, \nup413 (1994) 162.}
\ldf\villa{M.\ Villasante, \prv{D 45} (1992) 831.}
\ldf\feva{S.\ Ferrara and A.\ Van Proeyen, \cqg6 (1989) 124.}
\ldf\creva{E.\ Cremmer and A.\ Van Proeyen, \cqg2 (1985) 445.}
\ldf\cremme{E.\ Cremmer, S.\ Ferrara, L.\ Girardello
and A.\ Van Proeyen, \nup212 (1983) 413.}
\ldf\pvan{E.\ Cremmer, B.\ Julia, J.\ Scherk,\ S.\ Ferrara, L.\
Girardello and P.\ van Nieuwenhuizen,
\plt79B (1978) 231, \nup147 (1978) 105.}
\ldf\ntSUGRA{E.\ S.\ Fradkin, M.\ A.\ Vasiliev, Nuovo Cim. Lett.
{\bf 25} (1979) 79;
B.\ de Wit and J.\ W.\ van Holten, \nup155 (1979) 530;
B.\ de Wit, J.\ W.\ van Holten and A.\ Van Proeyen,
\nup167 (1980) 186.}
\ldf\fgkp{S.\ Ferrara, L.\ Girardello, C.\ Kounnas and M. Porrati,
\plt{B192} (1987) 368.}
\ldf\twomod{
A.\ Ceresole, R.\ D'Auria and T.\ Regge, \nup414 (1994) 517;
for a recent review, see A.\ Ceresole, R.\ D'Auria,
 S.\ Ferrara, W.\ Lerche,  J.\ Louis and T.\ Regge, \tit
Picard--Fuchs
Equations, Special Geometry and Target Space Duality | preprint
CERN-TH.7055/93,
POLFIS-TH.09/93, to appear on \tit Essays on Mirror Manifolds | vol.
II,
 S.\ T.\ Yau editor, International Press (1994).}
\ldf\veve{R.\ Dijkgraaf, E.\ Verlinde, H.\ Verlinde,
\cmp115 (1988) 649.}
\ldf\nara{K.\ Narain, \plt{169B} (1986) 41;
K.\ Narain, M.\ Sarmadi and E.\ Witten, \nup279 (1987)
369.}
\ldf\Shap{A.\ Shapere and F.\ Wilczek, \nup320 (1989) 669.}
\ldf\abk{I.\ Antoniadis, C.\ Bachas and C.\  Kounnas, \nup289 (1987)
87;
 H.\ Kawai, D.\ C.\ Lewellen and S.\ H.\ Tye, \nup288 (1987)1.}
\ldf\fepo{S.\ Ferrara and M.\ Porrati, \plt{B216} (1989) 289.}
\ldf\seib{N. Seiberg, \plt206B (1988) 75.}
\ldf\specs{N. Seiberg, \nup303 (1988) 206.}
\ldf\specst{S.~Ferrara and A.~Strominger, in \tit Strings
'89 | eds. R.~Arnowitt, R.~Bryan, M.J.~Duff, D.V.~Nanopoulos and
C.N.~Pope(World Scientific, 1989), p.~245;
A.\ Strominger, \cmp133 (1990) 163.}
\ldf\speccdf{L.\ Castellani, R.\ D'Auria and S.\ Ferrara, \plt241B
(1990)
57; \cqg1 (1990) 317.}
\ldf\specco{P.\ Candelas and X.\ de la Ossa, \nup355 (1991) 455;
P.\ Candelas, X.\ de la Ossa, P.\ S.\ Green and L.\ Parkes,
\plt258B (1991) 118; \nup359 (1991) 21. }
\ldf\specfl{S.\ Ferrara and J.\ Louis, \plt278B (1992) 240.}
\ldf\specere{A.\ Ceresole, R.\ D'Auria, S.\ Ferrara, W.\ Lerche and J.
Louis,
\ijmp8 (1993) 79}
\ldf\fors{
A.\ C.\ Cadavid and S. Ferrara, \plt267B (1991) 193;
W.\ Lerche, D.\ Smit and N.\ Warner, \nup372 (1992) 87.}
\ldf\filq{A.\ Font, L. Ibanez, D.\ Lust, F.\ Quevedo, \plt249B (1990)
35.}
\ldf\bate{A.\ Erd\'elyi, F.\ Obershettinger, W.\ Magnus and F. G.
Tricomi,
\tit Higher Transcendental Functions|, McGraw Hill,
New York, (1953).}
\ldf\nfour{
A.\ Sen, \nup388 (1992) 457 and \plt303B (1993) 22;
A.\ Sen and J.\ H.\ Schwarz, \nup411 (1994) 35; \plt312B (1993) 105.}
\ldf\dk{
M.\ J.\ Duff and R.\ R.\ Khuri, \nup411 (1994) 473.}
\ldf\gh{J.\ Gauntlett and J.\ A.\ Harvey, \tit S--Duality and the Spectrum
of Magnetic Monopoles in Heterotic String Theory | preprint EFI-94-36,
 hep-th/9407111.}
\ldf\ggpz{
L.\ Girardello, A.\ Giveon, M.\ Porrati and A.\ Zaffaroni,
\plt{B334} (1994) 331, hep-th/9406128.}
\ldf\tsd{For a review on target space duality see A.\ Giveon,
M.\ Porrati and E.\ Rabinovici, \prp244 (1994) 77, hep-th/9401139.}
\ldf\nvsov{A. Das, \prv{D15} (1977) 2805;
E. Cremmer, J. Scherk and S. Ferrara, \plt {68B} (1977)
234;
E. Cremmer and J. Scherk, \nup127 (1977) 259.}
\ldf\csf{E.\ Cremmer, J.\ Scherk and S.\ Ferrara, \plt{74B} (1978)
61.}
\ldf\susu{B.\ de Wit and A.\ Van Proeyen, \nup245 (1984) 89;
E.\ Cremmer, C.\ Kounnas, A.\ Van Proeyen, J.\ P.\ Derendinger, S.\
Ferrara,
B.\ De Wit and L.\ Girardello, \nup250 (1985) 385;
B.\ de Wit, P.\ G.\ Lauwers and A.\ Van Proeyen, \nup255 (1985) 569;
S.\ Cecotti, S.\ Ferrara and L.\ Girardello, \ijmp4 (1989) 2475.}
\ldf\fklz{S.\ Ferrara, C.\ Kounnas, D.\ Lust and F.\ Zwirner, \nup365
 (1991) 431.}
\ldf\fkdz{J.\ P.\ Derendinger, S.\ Ferrara, C.\ Kounnas
and F.\ Zwirner, \nup372 (1992) 145.}
\ldf\strNtwo{B.\ de Wit, J.W.\ van Holten and A.\ Van Proeyen,
\nup184 (1981) 77}
\ldf\freans{D. Anselmi and P. Fr\`e, \nup404 (1993) 288; and \tit
Gauged Hyperinstantons and Monopole equations | preprint
HUTP-94/A041, SISSA 182/94/EP, hep-th/9411205.}
\ldf\ewitt{ E.\ Witten, \tit Monopoles and Four--Manifolds |
 preprint IASSNS-HEP-94-96, hep-th/9411102.}
\ldf\Prso{M.\ K.\ Prasad and C.\ M.\ Sommerfield, \prl35
 (1975) 760;
E.\ B.\ Bogomol'nyi,  Sov.\ J.\ Nucl.\ Phys.\ $\us{24}$ (1976) 449;
E.\ Witten and D.\ Olive, \plt{78B} (1978) 97;
C.\ Montonen and D.\ Olive, \plt{72B} (1977) 117;
P.\ Goddard, J.\ Nuyts and D.\ Olive, \nup125 (1977) 1.}
\ldf\asen{A.\ Sen, \ijmp9 (1994) 3707, hep-th/9402002 and references
there in.}
\ldf\harliu{J. Harvey and J. Liu, \plt {B268} (1991)
40.}
\ldf\khu{R.R. Khuri, \plt{B259} (1991) 261;  \plt{B294} (1992) 325.}
\ldf\kaor{R. Kallosh and T. Ortin, \prv{D48} (1993) 742,
hep-th/9302109.}
\ldf\gibhulman{G.\ Gibbons and C.\ Hull, \plt{B109} (1982)
190; G.\ Gibbons and N.\ Manton, \nup274 (1986) 183.}
\ldf\censor{R.\ Kallosh, A.\ Linde, T.\
Ort\'{\i}n, A.\ Peet and A.\ Van Proeyen, \prv {D46} (1992) 5278.}
\ldf\HuTo{C.\ M.\ Hull and P.\ K.\ Townsend,
\tit Unity of Superstring Dualities | preprint
QMW 94-30, hep-th/9410167.}
\ldf\sgthree{S.\ Ferrara, C.\ Kounnas, \nup 328 (1989) 406;
S.\ Ferrara, P.\ Fre \ijmp 5 (1990) 989.}

\chap{Introduction}
Recently, proposals for the quantum moduli space of $N=2$
rigid Yang--Mills theories \seiwit\
 have been given in terms of particular classes of genus $r$
Riemann surfaces parametrized by $r$ complex moduli\KLTY
, $r$ being the rank
for the gauge group $G$ broken to $U(1)^r$  for generic values of the
moduli. The effective action for such theories, with terms up to two
derivatives, is described by $N=2$ supersymmetric lagrangians of $r$
abelian massless vector multiplets\susu, whose dynamics is encoded in
a holomorphic prepotential  $F(X^A)$, function of the moduli
coordinates
$X^A$ ($A=1,\ldots,r$).
According to Seiberg and Witten \seiwit\ this effective theory has
classical,
perturbative and non perturbative duality symmetries which reflect on
monodromy properties of certain holomorphic symplectic vectors
$(X^A,F_A(X))$,  eventually related to periods of holomorphic
one--forms\seiwit
$$
\omega=X^A \a_A +F_A \b^A \ ,
\eqn{\iuno}
$$
where $\a_A, \b^A$ is a basis for the $2r$ homology cycles of a genus
$r$
Riemann surface.
The Picard--Fuchs equations satisfied by the holomorphic vector
 one--form $U_i=(\del_i X^A,\del_i F_A)$ ($i=1,\ldots,r$)
 can be regarded as differential identities
for ``rigid special geometry'' \NOI.
To attach a particular algebraic curve to ``rigid special geometry" is
therefore
equivalent to exactly compute the holomorphic data $U_i$, and thus
to
exactly reconstruct the effective action for the self interaction of
the $r$
massless gauge multiplets  once the massive states, both perturbative
and
non perturbative, have been integrated out.
Indeed it is a virtue of $N=2$ supersymmetry that all the couplings
in the
effective Lagrangian, including $4$--fermion terms, can be computed
purely
in terms of the holomorphic data.
Quite remarkably the quantum monodromies dictate the monopole and
dyon spectrum
of the effective theory \doubref\seiwit\KLTY\ which turns out to be
``dual" to non--perturbative
instanton effects \AMAT\ in the original $G$--invariant microscopic
theory \doubref\freans\ewitt .

This paper considers several issues in order to extend the
approach
pursued in the rigid case to the more challenging case of coupling an
$N=2$
Yang--Mills theory to gravity. In particular we shall include in the
$N=2$
supergravity theory a dilaton--axion vector multiplet which is an
essential
ingredient to describe effective $N=2$ theories which come from the
low
energy limit of $N=2$ heterotic string theories in four dimensions
\fgkp.
Another ingredient is the extension of the ``classical monodromies" to
$N=2$
local supersymmetry. For rigid theories the classical metric is
essentially
the Cartan matrix of the group $G$ and the classical monodromies are
related to the Weyl group of the Cartan subalgebra of $G$ \KLTY . For
$N=2$
supergravity theories coming from $N=2$ heterotic strings, the
classical
metric of the moduli space of the pure gauge sector is based on the
homogeneous
space $ O(2,r)/O(2)\times O(r)$
\nrf{\specs\fepo}\refs{\susu{,}\fgkp{--}\fepo} and the
classical monodromies are related
to the $T$--duality group $ O(2,r;\IZ)$ which in particular is an
invariance
of the massive charged states\tsd .
This state of affair is quite analogous to the analysis performed by
Sen
and Schwarz\nfour\ for the $N=4$ heterotic string compactifications,
in which case
an exact quantum duality symmetry $ SL(2,\IZ)\times O(6,r;\IZ)$ was
conjectured \nrf{\dk\gh\ggpz\jhs}\refs{\nfour{--}\jhs}
and a resulting spectrum for BPS states with both electric and
magnetic states
was proposed.
In the $N=4$ theory the $SL(2,\IZ)\times O(6,r;\IZ)$ symmetry, using
general
arguments \doubref\FSZ\GAZU, has a natural embedding in
$Sp(2(6+r);\IZ)$, acting on the $6+r$
vector self--dual field strengths $\cF_{\mu\nu}^{+A}$ and their
``dual" defined
through
  $G_{+A}^{\mu\nu}\equiv -i{{\d \cL}\over{\d \cF^{+A}_{\mu\nu}}}$.
In generic $N=2$ theories, because of quantum corrections
\doubref\dive\seib ,
we do not expect
such factorized $S-T$ duality to occur anymore\NOI . Indeed this can be
argued
with a pure supersymmetry argument, related to the fact that once the
classical
moduli space $O(2,r)/O(2)\times O(r)$ is deformed by quantum
corrections, then
the factorized structure with the dilaton degrees of freedom is lost
and
a non trivial moduli space, mixing the $S$ and $T$ degrees of freedom
should
emerge. This result is in fact a consequence of a theorem on ``special
geometry" \doubref\feva\creva\
which asserts that the only factorized special manifolds are the
${{SU(1,1)}\over{U(1)}}\times{{O(2,r)}\over{O(2)\times O(r)}}$
series, which
precisely describe the ``classical moduli space" of $S-T$ moduli.
Because of the coupling to gravity, the symplectic structure and
identification
of periods, coming from special geometry, is also remarkably
different from rigid special geometry.
Indeed the interpretation of $(X^\L,F_\L),\ \L=0,1,\dots,r+1$ as
periods of
algebraic curves is no longer appropriate to genus $r$ Riemann
surfaces, as
it can be seen from the Picard--Fuchs
equations \doubref\specco\fors\
and from the form of the
metric $g_{i \jb} =- \del_i \del_\jb \log i (\bF_A X^A-\bX^A F_A)$ of
the
moduli space
\nrf{\specst\speccdf\DFF\specfl\specere}\refs{\specco{--}\specere} .
In fact special geometry is known to be appropriate to a particular
class
of  complex manifolds (Calabi--Yau manifolds or their mirrors)
and to describe the deformations of the complex structure\specco.
 It is therefore
tempting to argue that the quantum moduli space including $S-T$
duality
and its monodromies is related to  $3$--manifolds (or their mirrors)
 with $h_{(2,1)}=r+1$.

The paper is organized as follows:
In chapter $2$ we give a r\'esum\'e of rigid theories, also discussing
duality
for the fermionic sector and the physical significance of monodromies
and geometrical data, such as the holomorphic tensor $C_{ijk}$,
related to
the gaugino anomalous magnetic moment.
In chapter $3$ we describe in detail the coupling to gravity, the
extension of
duality to the fermionic sector and the existence of  symplectic
bases
which do not admit a prepotential function $F$, as it occurs in
certain
formulations of $N=2$ supergravities coming from $N=2$ heterotic
strings. The general form of duality transformations and symmetries
as they occur in $N=1$ locally supersymmetric Yang--Mills theories
coupled to matter is also described.
In chapter $4$ we use such a formulation
where all the perturbative duality symmetries become invariances of
the action. Then, we discuss the implementation of duality symmetries
in $N>2$ extended supergravities for the spectrum of dyonic states.
In chapter $5$ we analyze classical and quantum duality symmetries
and give generic formulae for the spectrum of the BPS states and the
``semiclassical formulae" when the non perturbative spectrum is
computed in terms of the ``classical periods". The explicit
expression for the $r=2$ case is given as  an example, and the special
occurrence of enhanced symmetry points is described. The paper
ends with some concluding remarks.

\chap{R\'esum\'e of rigid special geometry}
\section{Basics}

\ni
$N=2$ supersymmetric gauge theory on a group $G$ broken to $U(1)^r$,
with $r=rank\ G$, corresponds to a particular case
of the most general $N=1$ coupling of $r$ chiral multiplets
$(X^A,\chi^A)$
to $r$ $N=1$ abelian vector multiplets $(\cA_\mu^A, \l^A)$ in which
the \K potential $K$ and the holomorphic kinetic term function
$f_{AB}(X^A)$
are given by
$$\eqalign{
K &=i (\bF_A X^A-F_A \bX^A)\ \ , \ \ (F_A=\del_A F)\cr
f_{AB} &=\del_A\del_B F\equiv F_{AB}\cr}
\eqn\buno
$$
in terms of the single prepotential $F(X)$\susu. One can show that
the \K geometry is constrained because the Riemann tensor satisfies
the
identity \doubref\speccdf\NOI
$$
R_{A \bar B C \bar D}=-\del_A \del_C \del_P F~
 \del_{\bar B} \del_{\bar D} \del_{\bar Q} \bF~ g^{P \bar Q}\ ,
\eqn\bdue
$$
with
$$
g_{P \bar Q}=\del_P \del_{\bar Q} K= 2 ~\Im \del_P \del_Q F\ .
\eqn\btre
$$

The lagrangian has the form
$$
\eqalign{
\cL &= g_{A \bar B}\partial_\mu X^A \partial_\mu \bX^B+
 \ (g_{A\bar B} \l^{I A} \s^\mu\cD_\mu \bar\l^{\bar B }_I+
{\rm h.c.})\cr
{}~& +~ \Im (  F_{AB}\cF_{\mu\nu}^{-A}\cF_{\mu\nu}^{-B})
+\cL_{\rm Pauli}+\cL_{\rm 4-fermi}\ ,\cr}
\eqn\bquat
$$
where $A,B,\ldots$ run on the adjoint representation of the gauge
group
$G$,
$I=1,2$ and $\cF^{+A}_{\mu\nu}=\cF^A_{\mu\nu}-{i\over2}
\e_{\mu\nu\rho\s}\cF^{A\rho\s}$  (and $\cF^{-A}_{\mu\nu}=\bar
\cF^{+A}_{\mu\nu}$).
 As we shall see, also $\cL_{\rm Pauli}$
and $\cL_{\rm 4-Fermi}$  contain the function $F$ and its derivatives
up to
the fourth.

The previous formulation, derived from tensor calculus, is incomplete
because it is not coordinate covariant.
It is written in a particular coordinate system  (``special
coordinates'')
which is not uniquely selected. In fact, eq.\buno\ is left
invariant under particular coordinate changes of the $X^A \to \tilde
X^A$ with
some new function $\tilde F (\tilde X)$ described by

$$
\eqalign{
\tilde X^A(X) &= A^A_{\ B}X^B+B^{AB}F_B(X)+P^A\cr
\tilde F_A(\tilde X^A(X)) &= C_{AB} X^B +D_A^{\ B} F_B(X) +Q_A\ ,\cr
}\eqn\bcin
$$
where $\pmatrix{A&B\cr C&D\cr}$ is an $Sp(2r,\IR)$ matrix
$$
A^T C-C^T A=0\ \ ,\ \ B^T D- D^T B=0\ \ , \ \  A^T D-C^T B=\bigone\ ,
\eqn\bsei
$$
and $P^A, Q_A$ can be complex constants which
from now on will be set to zero.

It can be shown that a function $\tilde F$ exists such that \susu
$$
\tilde F_A={{\del \tilde F}\over{\del \tilde X^A}}\ ,
\eqn\bset
$$
provided the mapping $X^A\to \tilde X^A$ is invertible.

It is well known that  the equations of motion and the Bianchi
identities
\susu\FSZ\GAZU
$$
\eqalign{
\del^\mu \Im \cF^{-A}_{\mu\nu} &=0\ \ \ \ \ {\rm Bianchi\
identities}\cr
\del_\mu \Im G_{-A}^{\mu\nu} &=0\ \ \ \ \  {\rm Equations\  of\
motion}\cr
}\eqn\botto
$$
transform covariantly under \bcin\  (with $P^A=Q_A=0$), so that
$(\cF^{-A}_{\mu\nu},G_{-A}^{\mu\nu})$ is a symplectic vector. Here,
$G_{-A}^{\mu\nu}\equiv i{{\d \cL}\over{\d \cF^{-A}_{\mu\nu}}}=
\bar\cN_{AB} \cF^{-B}_{\mu\nu}+{\rm fermionic}\ {\rm terms}$,
where  we have set $F_{AB}=\bar\cN_{AB}$
 in order to unify the notations to the
gravitational case\susu. The transformations \bcin\ leave
invariant the whole lagrangian but the vector kinetic term.
 Indeed, neglecting for the moment
fermion terms
(see section 2.2) and setting for simplicity
$\cF^{-A}_{\mu\nu}=\cF^A$ and
$G_{-A}^{\mu\nu}=G_A$ the vector kinetic lagrangian transforms
as follows
$$
\eqalign{
\Im  \cF^A {\bar\cN}_{AB}  \cF^B &\to
\Im \tilde\cF^A \tilde G_A =\cr
 &= \Im (\cF^A G_A +2 \cF^A(C^T B)_A^{\ B} G_B +\cr
 &+ \cF^A(C^T A)_{AB} \cF^B+G_A(D^T B)^{AB} G_{B} )\ .\cr
}\eqn\bdieci
$$
If $C=B=0$ the lagrangian is  invariant. If $C\neq 0, B=0$ it is
 invariant up to a four--divergence.
In presence of a topologically
non--trivial
$\cF^{-A}_{\mu\nu}$ background,   $(C^T A)_{AB}\int \Im
\cF^{-A}_{\mu\nu}
 \cF^{-B}_{\mu\nu}
\neq 0$, one sees that in the quantum theory
 duality transformations must be integral valued in
$Sp(2r,\IZ)$\seiwit\ and transformations with $B=0$ will be called
perturbative duality
transformations.

If $B\neq 0$  the lagrangian is not invariant. As it is well known,
 then the duality transformation is
only a symmetry of the equations of motion and not of the lagrangian.

Since
 $\tilde G^{\mu\nu}_{-A}=\tilde{\bar\cN}_{AB}{\tilde\cF}^{-B}_{\mu\nu
}$
one also has
$$
\tilde \cN=(C+D \cN)(A+B \cN)^{-1}\ .
\eqn\bundi
$$

A duality transformation will be a symmetry of the theory if
$\tilde\cN(\tilde X)=\cN(\tilde X)$, which implies $\tilde F(\tilde X)
= F(\tilde X)$.

Note that $B\neq 0$ means that the coupling constant $\tilde\cN$ is
inverted and symmetry
 transformations with $B\neq 0$ will be called quantum
non perturbative duality symmetries.

 The perturbative duality rotations are of the form
$$
\pmatrix{A&0\cr C& (A^T)^{-1}\cr}\ \ , \ A\subset GL(r)\  \  , \ A^T
C
 \ \ {\rm symmetric}\ .
\eqn\bdodi
$$
In rigid supersymmetry the tree level symmetries are of the form
$\pmatrix{A & 0\cr 0 & (A^T)^{-1}\cr}$ while the quantum perturbative
 monodromy introduces a $C\neq 0$.

The general form of the central charge for BPS states in a generic
$N=2$ rigid theory is given by \seiwit
$$
\mid Z\mid=M=\mid n^A_{(m)} F_A-n^{(e)}_A X^A \mid\ ,
\eqn\centra
$$
where $n_{(m)}^A\ , n^{(e)}_A$ denote the values of magnetic
 and electric charges of the state of mass $M$. The above expression
is manifestly symplectic covariant provided the vector
$(n_{(m)}^A,n^{(e)}_A)$ is also transformed under $Sp(2r;\IZ)$. This
equation shows again that a duality symmetry can only be a
(perturbative) symmetry if $B=0$, otherwise the vector subspace
with $n^A_{(m)}=0$ cannot be left invariant.

If the original unbroken  gauge group is $G=SU(r+1)$, then $A\in$
Weyl
group and $A^T C$ is the Cartan matrix $<\a_i |\a_j>$
of $SU(r+1)$\KLTY.

Eq. \bundi\ shows that $A+B{\cal N}$ has to be invertible in order
that the new tensor $\tilde{\cal N}$ exists. This is insured by the
positive definiteness of $\Im {\cal N}$, which is the kinetic
matrix. Here $A+B{\cal N} = \partial \tilde X/\partial X$, so this
implies the invertibility of the mapping $ X\to \tilde X$. As
explained in \bset, this then also implies the existence of $\tilde
F$. We will see that in local supersymmetry ${\cal N}_{AB}\neq \bar
F_{AB} $, so that the existence of $\tilde F$ is not equivalent to
the invertibility of $\Im {\cal N}$, and $\tilde F$ not always exists.

Special coordinates do not give a coordinate independent description
of the effective action. A coordinate independent description
is obtained by introducing a holomorphic symplectic bundle
$V=(X^A(z),F_A(z))$
and holomorphic $(1,0)$ forms on the \K manifold\doubref\NOI\seiwit
$$
U_i\equiv \del_i V=(\del_i X^A,\del_i F_A) \qquad{\rm with }\
i=1,\ldots,r \ .
\eqn\btredi
$$
 In rigid special geometry the  $U_i$ satisfy the constraints\NOI
$$
\eqalign{
\cD_i U_j &=i C_{ijk} g^{k\bar l} \bar U_{\bar l}\cr
\del_i \bar U_{\jb} &=0\ .\cr}
\eqn\btred
$$
Taking then the metric
$$
\eqalign{
g_{i\jb} &=\del_i\del_\jb K =i(\del_\jb \bF_A \del_i X^A-\del _\jb
\bX^A\del_i F_A)\cr
&=i \del_i X^A\del_\jb\bX^B
(\cN_{AB}-\bar\cN_{AB})\ ,\cr}
\eqn\bquindi
$$
where we used
$$
\del_\ib\bF_A=\cN_{AB}\del_\ib\bX^B\ ,
\eqn\bquattor
$$
one may derive the tensor $C_{ijk}$
$$
\eqalign{
C_{ikp} &=\del_i X^A \cD_k \del_p F_A-\del_i F_A \cD_k \del_p X^A
\cr
\ &= \del_i X^B(\del_k \del_p F_B-\del_k \del_p X^A \bar\cN_{AB})\ .
\cr }\eqn\bdicias
$$
The integrability conditions on \btred\ yields
$$
R_{i\bar j k \bar l}=- C_{ikp} \bar C_{\bar j \bar l \bar p} g^{p
\bar p}\ .
\eqn\bsedi
$$

The Bianchi identities of \bsedi\ also imply that $C_{ijk}$ is a
holomorphic
completely symmetric tensor obeying $\cD_{[i} C_{j]kl}=0$.

Note that from \bdicias\ it also follows
$$
C_{ijk}=\del_i X^A \del_j X^B \del_k X^C \del_A\del_B\del_C F\ ,
\eqn\extra
$$
which in special coordinates reduces to
$$
C_{ABC}= \del_A\del_B\del_C F\ .
\eqn\basta
$$

\section{Symplectic transformations in the fermionic sector}

In the total supersymmetric action, the vectors also couple to
fermions by terms linear in the field strength.
We will first give the general features of the formulation of
symplectic transformations in the presence of a fermionic sector,
which could even be non--supersymmetric. Afterwards, we will
specify the formulae for generic fermionic terms which we encounter
in $N=2$ lagrangians.

The general form of the Lagrangian, deleting terms which are by
themselves symplectic invariant, is
$$
{\cal L}=-\ft i2 \bar {\cal N}_{AB}\cF^{- A\mu\nu}  \cF^{-B}_{\mu\nu}
-i \cF^{-A\mu\nu} H^-_{A\mu\nu}
+c.c. +{\cal L}_{4f}\ ,  \eqn\LagV
$$
where $H^-_{A\mu\nu}$ are quadratic in the fermions,
and ${\cal L}_{4f} $ are the quartic terms in fermions. Then
$$
G^-_{A\mu\nu}\equiv i{\delta{\cal L}\over \delta \cF^{-A\mu\nu}}
=\bar {\cal N}_{AB} \cF^{-B}_{\mu\nu}+H^-_{A\mu\nu}=G^-_{bA\mu\nu}
+H^-_{A\mu\nu}\ . \eqn\Gfull
$$
As argued  in ref \FSZ,
the point where the equations of motions \botto\ are satisfied is an
invariant point.
Thus,  the first term of
the action is (omitting the obvious $A$ indices)
$$ \eqalign{
{\cal L}_V&\equiv -\ft i2 \bar {\cal N}\cF^-_{\mu\nu}\cF^{-\mu\nu}+c.c.
\cr
&=  -\ft i2 G^-_{b\mu\nu} \cF^{\mu\nu}+c.c. \cr
&=  i \partial^\mu  G^-_{b\mu\nu}  A^\nu +c.c. \cr
&= -i   \partial^\mu H^-_{\mu\nu} A^\nu +c.c.
-2\partial^\mu \Im G^-_{\mu\nu}A^\nu\cr
&= \ft i 2 H^-_{\mu\nu} \cF^{-\mu\nu}  +c.c.
-2\partial^\mu \Im G^-_{\mu\nu}A^\nu\ .
 } \eqn\calLV $$
Therefore
$$
\left.{\cal L}\right|_{{\delta{\cal L}\over \delta{\cal A}}=0}=
-\ft i 2 H^-_{\mu\nu} \cF^{-\mu\nu}  +c.c.+{\cal L}_{4f}
\equiv {\cal L}_{inv}\  ,\eqn\Linv
 $$
which should thus be invariant. The Lagrangian \LagV\ is then
$$
{\cal L}=-\ft i2 \cF^{-A\mu\nu} G^-_{A\mu\nu} + c.c.
+ {\cal L}_{inv}\ .\eqn\Lcompl
$$

Now we suppose $H^-_{A\mu\nu}$ to be of the form
$$
H^-_{A\mu\nu} =
\left(P_{Aa} -\bar {\cal N}_{AB} Q_a^B\right) \cT^{-a}_{\mu\nu} \ ,
\eqn\elegant
$$
where $a$ denotes a new index, whose meaning depends on the
model. $\cT^{-a}_{\mu\nu}$ is a tensor not transforming under the
symplectic group. Then
$$ \eqalign{
{\cal L}_{inv}&= -\ft i 2  \cF^{-A\mu\nu}
\left(P_{Aa} -\bar {\cal N}_{AB}
Q_a^B\right) \cT^{-a}_{\mu\nu} +c.c.+{\cal L}_{4f} \cr
&=  -\ft i 2 \left(  \cF^{-A\mu\nu} P_{Aa} - G_{bA}^{-\mu\nu}
Q^A_a\right) \cT^{-a}_{\mu\nu} +c.c.+{\cal L}_{4f} \ .
 } \eqn\Linvexpl $$
Invariance of  ${\cal L}_{inv}$ is then guaranteed if $(
Q^A, P_A)$ is a symplectic vector, and $ {\cal L}_{4f} $
is constructed as the completion of $G_b$ to $G$ in the above
formula (plus possible completely invariant terms). These completions
are thus
$$
{\cal L}_{4f} =\ft i 2  H^{-\mu\nu}_A Q^A_a  \cT^{-a}_{\mu\nu} +c.c.+
{\rm invariant\ terms}\ .         \eqn\Lfourf
$$

\section{Fermions in $N=2$ rigid Yang--Mills theory}

The coordinate independent description of fermions is given by $SU(2)$
doublets
 $(\l^{iI},\l^\ib_I)$ where upper and lower $SU(2)$ indices
$I$ mean positive and negative chiralities
respectively \susu\speccdf\DFF.
As such the spinors are symplectic invariant and contravariant
world vector fields.
The antiselfdual field strength $\cF^{-A}_{\a\b}$ and positive
chiralities spinors are in the same $N=2$ multiplet, which is, in two
component spinor notation,\foot{$\cF^{-A}_{\a\b}$ is
$\sigma^{\mu\nu}_{\alpha\beta}\cF^{-A}_{\mu\nu}$.}
$$
(X^A,\del_i X^A \l^{iI}_\a, \cF^{-A}_{\a\b})\ ,
\eqn\funo
$$
with $\a,\b \in SL(2,\IC)$.

In our application of \elegant\ only $\cT$ is dependent on the
fermions $\lambda^{iI}$, while $P$ and $Q$ depend on the scalars
$X^A$. The index $a$ is now replaced by $\ib$, and we have
$$ \eqalign{
Q_\ib^A&=\partial_\ib \bar X^A\ ;\qquad
P_{A\ib }= \partial_\ib \bar F_A\cr
\cT^\ib_{\alpha\beta}&=k g^{\ib j} C_{jkp}\lambda^{kI}_\alpha
\lambda^{pJ}_\beta\epsilon_{IJ} \ ,
 }\eqn\defQPT $$
where $k$ is a constant to be determined by supersymmetry. Then
$$
H_{-A}^{\a\b}= k\del_\ib\bX^B(\cN_{BA}-\bar
\cN_{BA})g^{\ib j}C_{jkp}\l^{\a k I}\l^{\b p J}\e_{IJ}\ .
\eqn\fcin
$$
This yields
$$\eqalign{
\cL_{\rm Pauli}=&-i({\cal N}-\bar {\cal N})_{AB} \del_\ib\bX^A
 \cT^\ib_{\alpha\beta}   \cF^{B\alpha\beta}+c.c.\cr
\cL_{\rm 4f}=&\ft i2\del_\ib\bX^A\del_{\jb}\bX^B
(\bar\cN_{AB}-\cN_{AB}) \cT^\ib_{\alpha\beta}   \cT^{\jb\alpha\beta}
+c.c.+\ {\rm invariant\ terms}\ , \cr}
\eqn\fsei
$$
in agreement with Cremmer et al. \cremme .

In special coordinates, setting
$\l^{i 1}_\a=\chi^i_\a,\  \l^{i2}_\a=\lambda^i_\a$,
 the Pauli term reduces to
$$
\cL_{\rm Pauli}=-k\  \del_A\del_B\del_C
F(\chi^A_\a\l^B_\b-\l_\a^A\chi^B_\b)
\cF^{-C\a\b } +c.c.\ ,
\eqn\fsette
$$
in agreement with the standard $N=1$ supersymmetric action with
$f_{AB}=F_{AB}$ \cremme.
We see from  \fsei\
 that in rigid supersymmetry the physical meaning of $C_{ijk}$
is that of an anomalous magnetic moment. Note that $C_{ijk}$ vanishes
at tree--level and it is $\sim {1\over{<X>}}$ at one loop-level as
it must be \refs\dive\seib\seiwit. It is obviously singular at $<X>=0$.
In the $SU(2)$ quantum theory \seiwit, the $SU(2)$ symmetry is not
restored at $X=0$, and then one rather expects
such terms to behave as ${{c_0}\over{\L}}$ where $c_0$ is a
dimensionless number.
The vanishing
at tree-level of both Pauli terms and the corresponding four fermions
terms
is consistent with renormalizability arguments.

The other fermionic terms which are already duality invariant read
$$
\l^{iI}_\a \l^{kJ}_\b \e^{\a\b} \bar\l^\jb_{\dot\a I}
\bar\l^{\bar l}_{\dot\b J} \e^{\dot\a\dot\b}
R_{i\jb k \bar l}
\eqn\fotto
$$
and
$$
\cD_i C_{jlm} \l^{iI}_\a \l^{jK}_\b \e^{\a\b}
\l^{lJ}_\c \l^{mL}_\d \e^{\c\d} \e_{IJ} \e_{KL}\ .
\eqn\fnove
$$

Note that, because of eq. \bsedi,  all couplings in the lagrangian
are
expressed through the tensors $C_{ijk}$.

{}From a tensor calculus point of view, all quartic terms but the
last come from the equations of motion
of the $Y^i_{IJ}$ auxiliary field triplet\susu.

\section{Positivity and monodromies}

Let us consider a submanifold $\cM_r$ of the moduli space of a
Riemann surface
of genus $r$ such that its tangent space is isomorphic to the Hodge
bundle.
In particular the dimension of $\cM_r$ is equal to the genus $r$ of
the Riemann surface
${\cal C}_r$\foot{We are aware of the fact that to find an intrinsic
 characterization of such an algebraic locus is far from obvious. We
thank D. Dubrovin, D. Franco, P. Fr\'e and C. Reina
for clarifying discussions on this point.}.
In this case, decomposing an abelian differential
in terms of the $2r$ harmonic forms dual to the canonical basis of
cycles,
we have
$$\eqalign{
&\omega=X^A(z^i)\a_A+F_A(z^i)\b^A\ \ \  \ A,i=1,\dots,r\cr
&\int\a_A\wedge \b^B=\d_A^{\ B}\ \
\ , \ \int\a_A\wedge\a_B=\int\b_A\wedge\b_B=0\ ,\cr
}
\eqn\puno$$
where $z^i$ are coordinates on the moduli space submanifold, and
$$
\del_i \omega =\del_i X^A\a_A+\del_i F_A\b^A \ .
\eqn\pdue
$$
Then the metric, given by the norm
$$
g_{\ib j}=i\int \del_i\omega\wedge\del_\jb \bar\omega
=i\del_i\del_\jb\int
\omega\wedge\bar\omega
\eqn\ptre
$$
is manifestly positive. Using eqs. \puno , \pdue\  we find
$$
g_{i \jb} =i \del_i \del_\jb (\bF_A X^A-\bX^A F_A)
$$
which coincides with the metric of $N=2$ rigid special geometry
\bquindi\ \doubref\seiwit\NOI.

Formula \pdue\ implies by supersymmetry
a similar expansion for the full multiplet \funo . For the upper
component $\cF^{-A}_{\mu\nu}$ we get a self dual three form
$$
w=\cF^A\a_A+G_A\b^A
\eqn\pcin
$$
on $\IR_4\times \cC_r$
when \botto\ hold.
We observe that an $N=2$, $4D$ abelian vector multiplet can be
obtained from dimensional reduction from six dimensions either of
a vector multiplet or of
a tensor multiplet containing a self--dual field strength.
This remarkable
coincidence
 actually suggests a physical
picture for the
characterization of this subclass $\cC_r$  of Riemann surfaces.
 Namely, they should appear in the compactification
  on $\IR_4\times \cC_r$ of $N=1$ six--dimensional theory
of a self interacting  tensor multiplet.

As shown in ref. \NOI,  the Picard--Fuchs equations for
$\cC_r$ have a general form dictated by the differential constraints of
rigid special geometry.
A general proposal for $\cC_r$ has been given in \KLTY\ and can be
used to
write down the Picard--Fuchs equations for the periods and to
determine
their monodromies. Such proposal can be checked by comparing the
explicit form
of the Picard--Fuchs equations with their general form given by rigid
special
geometry.

In the one parameter case ($G=SU(2)$), where $\cC_1$ is given by the
elliptic
 curve of
ref. \seiwit, the special geometry  equations reduce to
one ordinary second order equation
$$
({{d}\over{dz}}+\hat\Gamma)C^{-1}({{d}\over{dz}}-\hat\Gamma) U=0
\eqn\pset
$$
where $\hat\Gamma={{d}\over{dz}} \log e$, $e={{d X}\over{d z}}$ and
$C$ is the 3--tensor appearing in \btred. This
agrees with the Picard--Fuchs equations derived  from $\cC_1$.
The general solution of this equation is\NOI
$$
U=(e,e{{d^2 F}\over{dX^2}})\ \ \ ,
\eqn\pott
$$
with $\tau={{d^2 F}\over{d X^2}}$ being the uniformizing
variable
for which the differential equation reduces to
${{d^2}\over{d\tau^2}}(\ )=0$.

\chap{Coupling to gravity}
\section{Special geometry and symplectic transformations}

The coupling to gravity modifies the constraints of
 rigid special geometry
because of the introduction of a $U(1)$ connection due to the $U(1)$
\K--Hodge structure of moduli space.
For
 $n$ vector multiplets one introduces $2(n+1)$ covariantly
holomorphic
sections \refs{\speccdf{,}\specco{,}\DFF{,}\specere}
$$
V=(L^\L, M_\L)\ \ \ (\L=0,\ldots,n)\ ,
\eqn\cuno
$$
where $0$ is the graviphoton index.

The new differential constraints of special geometry are
$$
\eqalign{
U_i &\equiv (\cD_i L^\L,\cD_i M_\L)=(f^\L_i,h_{i\L})\cr
\cD_i U_j &= i C_{ijk} g^{k \bar l} \bar U_{\bar l}\cr
\cD_i \bar U_{\jb} &= g_{i\jb} \bar V\cr
\cD_i \bar V &= 0\ ,\cr}
\eqn\cdue
$$
where now $\cD_i$ is the covariant derivative with  respect to the
usual Levi-Civita connection and the \K connection $\del_i K$. That
is, under $K\to K+f+\bar f$ a generic field $\psi^i$ which under
$U(1)$
transforms as $
\psi^i\to e^{-({p\over2}f+{{\bar p}\over2}\bar f )}\psi^i$ has the
following covariant derivative
$$
\cD_i\psi^j=\del_i \psi^j +\G^j_{ik} \psi^k +{p\over2}\del_i K \psi^j
\ ,\eqn\ctre
$$
and analogously for $\cD_{\ib}$ with $p\to \bar p$. This $U(1)$ is
related to the $U(1)$ in the $N=2$ superconformal group, and the
weights for all the fields were determined in \strNtwo\ ($\bar p=c$).
In our notations, $(L^\L, M_\L)$ have been given conventionally
weights $p=-\bar p=1$.

Since $L^\L,M_\L$ are covariantly holomorphic, it is convenient to
introduce holomorphic sections $X^\L=e^{-K/2} L^\L $,
$F_\L=e^{-K/2}M_\L$.

The \K potential is fixed by
the condition\susu\speccdf\
$$
i(\bar L^\L M_\L-L^\L\bar M_\L)=1
\eqn\csei
$$
to be
$$
K=- \log i (\bX^\L F_\L-X^\L\bar F_\L)\ .
\eqn\csette
$$

As it is well known\susu\WVP, the differential constraints \cdue\
can in general
be solved in terms of a holomorphic function homogeneous of degree
two
$F(X)$. However, as we will see in the sequel, there exist particular
symplectic sections for which such prepotential $F$ does not exist.
In
particular this is the case appearing in the effective theory of
the $N=2$
heterotic string. For this reason it is convenient to have the
fundamental
formulas of special geometry written in a way independent of the
existence of $F$.

First of all we note that quite generally we may write
$$
M_\Lambda= {\cal N}_{\Lambda\Sigma}L^\Sigma \ ;\qquad
h_{\L i }=\bar \cN _{\L\Sigma}f^\Sigma_i\ .
\eqn\cqua
$$

{}From \cqua\ we can define the two $(n+1)\times (n+1)$ matrices
$$
h_{\L \bar I}=(h_{\L\bar 0}\equiv M_\L ,h_{\L \ib})\ \ ,
\ \ f^\L_{\bar I}=(f^\L_{\bar 0}\equiv L^\L , f^\L_{\ib})
\eqn\wspe
$$
to obtain an explicit expression for $\cN_{\L\S}$ in terms of
$(L^\L,M_\L)$ as
$$
\cN_{\L\S}=h_{\L\bar I}(f^{-1})^{\bar I}_\S\ .
\eqn\sergione
$$
Note that $h_{\L\bar I},f^\S_{\bar I}$ are invertible matrices and
the above expression implies the transformation law \bundi.

When $F$ exists, $\cN_{\L\S}$ has the form\susu\DFF
$$
\cN_{\L\Sigma}=\bar F_{\L\Sigma} +2i {{(\Im F_{\L\G})( \Im
F_{\Sigma\Pi})L^\G
L^\Pi}\over{(\Im F_{\Xi\Omega})\ L^\Xi L^\Omega}}\ ,
\eqn\ccin
$$
which turns out to be the coupling matrix appearing in the kinetic
term of the vector fields. However, as we show below, \cqua\ are
symplectic covariant and therefore they always hold even in some
specific coordinate system in which $F$ does not exist.

In the same way as in the rigid case, from eqs. \cdue\ and \csei\ we
find
$$
g_{i\jb}=i(f^\L_i\bar h_{\jb\L}-h_{ i\L}\bar f^\L_\jb )=
i( \cN_{\L\Sigma}-\bar\cN_{\L\Sigma})f^\L_i f^\Sigma_\jb
\eqn\cotto
$$
$$
C_{ijk}=f^\L_i \cD_j h_{k\L}-h_{ i\L}\cD_j f^\L_k=
f^\L_i\del_j\bar \cN_{\L\Sigma}f^\Sigma_k \ ,
\eqn\cnove
$$
which are symplectic invariant. (Note that $\cN_{\L\Sigma}$ has zero
\K weight).

Furthermore, the integrability conditions \cdue\
give\susu\speccdf\specst\specco\specere\DFF
$$
R_{i\jb l\bar k}=g_{i\jb} g_{l\bar k}+g_{i\bar
k}g_{l\jb}-C_{ilp}C_{\jb
\bar k \bar p} g^{p\bar p} \ ,
\eqn\cdiec
$$
replacing eq. \bsei.

Here $C_{ilp}$ is a covariantly holomorphic tensor of weight
$p=-\bar p=2$,
$$
\cD_{\bar l}C_{ijk}=\del_{\bar l} C_{ijk}-\del_{\bar l} K C_{ijk}=0
\ , \eqn\cundi
$$
which implies $\del_{\bar l} W_{ijk}=0$ with  $C_{ijk}=e^K
W_{ijk}$.

Some additional consequences of the previous formulae are the
following: from $\cD_i F_\L= \bar\cN_{\L\Sigma}\cD_i X^\Sigma$, applying
$\cD_\jb$
to both sides we also find
$$
\cD_\jb \cD_i F_\L=\del_\jb \bar\cN_{\L\Sigma} \cD_i X^\Sigma
+\bar\cN_{\L\Sigma}\cD_\jb \cD_iX^\Sigma  \ ,
\eqn\tdieci
$$
which implies, using the third line of \cdue,
$$
(F_\L-\bar\cN_{\L\Sigma} X^\Sigma )g_{i\jb}=\del_\jb
\bar\cN_{\L\Sigma}\cD_i X^\Sigma \ .
\eqn\tundi
$$
Note that the left--hand side of \tundi\ defines the graviphoton
projector
$$
T_\L= M_\L-\bar\cN_{\L\S}L^\S\ .
\eqn\ttundi
$$
{}From  the first of equations  \cqua\ it also follows that
$$
\del_\ib\cN_{\L\S}L^\S=0\ \ \ ,\ \ \ h_{i\L}=\cN_{\L\S}f_i^\S+
\del_i \cN_{\L\S} L^\S
\eqn\ttdodi
$$
and therefore
$$
\del_i \cN_{\L\S} L^\S=(\bar\cN_{\L\S}-\cN_{\L\S})f_i^\S
\eqn\sergg
$$
by contraction with $f_\jb^\L$ we get
$$
f^\L_\jb \del_i \cN_{\L\S} L^\S= i g_{i\jb} \ .
\eqn\serrg
$$
Taking the complex conjugate of \serrg\ and using \tundi\
it follows that
$$
T_\L \bar L^\L=-i\ .
\eqn\sserg
$$
which is nothing but \csei. An alternative form for the \K potential
is
$$
K=-\log\ i (\cN_{\L\S}-\bar\cN_{\L\S}) X^\L \bX^\S\ .
\eqn\aserg
$$

Duality transformations are now in $Sp(2n+2,\IZ)$ and act on
$X^\L,F_\L$ as
in the rigid case.
The symplectic action on $(L^\L,M_\L)$ (or $(X^\L,F_\L)$) is
$$
\pmatrix{L\cr M}'=\pmatrix{A&B\cr C&D \cr }\pmatrix{L\cr
M\cr}=\cS\pmatrix{L\cr
M\cr} \ \ \ \cS\in Sp(2n+2,\IZ)\ .
\eqn\cdodi
$$
Then it follows, because of eq. \cdue\ and \cqua,
$$
\pmatrix{f^\L_i\cr h_{i\L }}'
=\pmatrix{A&B\bar\cN\cr C&D\bar \cN \cr }\pmatrix{f_i^\L\cr
f_i^\L\cr} \ ,
\eqn\ctred
$$
which implies again \bundi.
These two transformations laws imply the covariance of \cqua .

The symplectic action on $\cF^{+\L}_{\mu\nu},G_{+\L}^{\mu\nu}$
 is the same as on $(L^\L, M_\L)$, so
eq. \botto\ is unchanged. Therefore the discussion of the previous
section
on perturbative and non perturbative duality transformations in the
rigid case remains unchanged when gravity is turned on.

When the sections $(X^\L,F_\L)$ are chosen in such a way that a
function
$F$ exists\foot{A r\'esum\'e of the duality transformations for this
case, including the supergravity corrections has been given in
appendix C of \WVP.},
from \csei\ and the degree two homogeneity of $F$ it
follows that \speccdf\DFF
$$
\Im\! F_{\L\S}\  L^\L\bar f^\S_\ib=0 \ ,
\eqn\iden
$$
so that the second of eq. \cqua\ becomes
$h_{i\L}=F_{\L\S}f^{\S i}$. Furthermore
from \cnove\ and \iden\ it also follows
$$
e^{K/2}C_{ijk}=f^\L_i f^\G_j f^\S_k F_{\L\G\S}\ .
\eqn\ricca
$$
By the same token, we have
$$
\pmatrix{f^\L_i\cr h_{i\L }}'
=\pmatrix{A&B\cF\cr C&D\cF \cr }\pmatrix{f_i^\L\cr f_i^\L\cr}\ ,
\eqn\ricco
$$
where $\cF=F_{\L\S}$. Note that in these cases
$$
\eqalign{
2\tilde F(\tilde X) &=\tilde F_\L\tilde X^\L=\cr
{}~&2F+2X^\L(C^TB)^\S_\L F_\S+X^\L (C^TA)_{\L\S}
X^\S+F_\L (D^TB)^{\L\S}F_\S\ .\cr}
\eqn\dcin
$$

Note also that the homogeneity of $F$ implies
$$
\tilde X=(A+B\cF) X \ ,
\eqn\cquin
$$
where $\cF=F_{\L\Sigma}$
and
$$
\tilde F=(C+D\cF)X\ .
\eqn\csedi
$$

Special coordinates in supergravity are defined by $t^\L=X^\L/X^0$
since
we now have a set of $n+1$ homogeneous coordinates.
If we assume that $\cD_i({{X^\L}\over{X^0}})$
is an invertible matrix, then we may choose a frame for which
$\del_i ({{X^\L}\over{X^0}})=\delta_i^\L$. This is possible only if
$X^\L$
are unconstrained variables and so $F_\L=F_\L(X)$, which implies
$F_\L=\del_\L
F(X)$ with $F$ homogeneous of degree 2.

We now discuss the possible non-existence of $F(X)$.
If we start with some special coordinates $X^\L,F_\L(X)$, it is
possible
that in the new basis the $\tilde X^\L$ are not good special
coordinates in
the sense that the mapping $X\to\tilde X$ is not invertible. This
happens
whenever the $(n+1)\times (n+1)$
 matrix $A+B\cF$ is not invertible (its determinant
vanishes). This does not mean that $\tilde X,\tilde F$ are not good
symplectic
sections since the symplectic matrix $\cS=\pmatrix{A&B\cr C&D\cr}$ is
always
invertible. It simply means that $\tilde F_\L\neq\tilde F_\L(\tilde
X)$ and
therefore a prepotential $\tilde F (\tilde X)$ does not exist.
However our
formulation of special geometry never explicitly used the fact that
$F_\L$
be a functional of the $X$'s and indeed the quantities $(X^\L,F_\L)$,
 $(f_i^\L,h_{i\L})$, $\cN_{\L\Sigma}$ and $C_{ijk},g_{i\jb}$ are well
 defined for any choice
of the symplectic sections $(X^\L,F_\L)$ since they are symplectic
invariant or covariant.
For example, to compute the ``gauge coupling'' $\tilde
\cN$ in such a basis $(\tilde X^\L,\tilde F_\L)$ one uses the formula
$$
\tilde\cN(\tilde X,\tilde F)=(C+D\cN (X))(A+B\cN(X))^{-1}\ ,
\eqn\cdicia
$$
and expresses the $X=X(\tilde X,\tilde F)$ by using the fact that the
symplectic
mapping can be inverted. All other quantities can be computed in this
way.

We will see the relevance of this observation in the sequel, while
discussing
low energy effective action of $N=2$ heterotic string.
A simple example is the following. Consider $F=iX^0X^1$, leading to
$$
{\cal N}=\pmatrix{i{X^1\over X^0}&0\cr 0& i{X^0\over X^1}\cr}\ .
\eqn\calNex
$$
This appears in the $N=2$ reduction of pure $N=4$ supergravity in the
so--called $SO(4)$ formulation \nvsov. Consider now the
symplectic mapping defined by
$$
A=D=\pmatrix{1&0\cr 0&0\cr}\ ;\qquad C=-B=\pmatrix{0&0\cr 0&1\cr}\ .
\eqn\ABCDex
$$
Then the transformation is
$$\eqalign{
\tilde X^0=X^0 \qquad& \tilde X^1=-F_1\cr
\tilde F_0=F_0 \qquad& \tilde F_1= X^1\ .\cr}
\eqn\transex
$$
Using in the first line $F_1=iX^0$ would lead to a non--invertible
mapping $X\to \tilde X$, and using \dcin\ would lead to $\tilde F=0$.
One observes also that $A+B{\cal F}$ is
non--invertible. However, $A+B{\cal N}$ is invertible, and one
obtains $\tilde \cN = i X^1(X^0)^{-1}\bigone =
i\tilde F_1 (\tilde X^0)^{-1}\bigone$. This form appears in
the $N=2$ reduction of the $SU(4)$ formulation of pure $N=4$
supergravity  \csf. These two forms of the $N=2$ reduced action and
the duality transformation
have been studied in \censor\ to relate electric and magnetic charges
of black holes.
\section{The fermionic sector}
As far as the fermions are concerned, the vector $N=2$
multiplet is now
$$
(L^\L, f^\L_i \l^{iI}_\alpha,  \cF^{-\L}_{\a\b})\ .
\eqn\multi
$$
The tensor $\cT^\ib_{\alpha\beta}$ is still the same as in \defQPT, and
$$
Q_\ib^\Lambda={\cal D}_\ib \bar  L^\Lambda \ ;\qquad
P_{\Lambda\ib}= {\cal D}_\ib \bar  M_\Lambda\ .
\eqn\culti
$$
Correspondingly, the gaugino Pauli terms have  the  form
$$
i (\cD_\ib \bar L^\L G^{\a\b}_{b-\L}-\cD_\ib \bar M_\L
\cF^{-\L\a\b}) \cT^\ib_{\alpha\beta} \ ,
\eqn\serg
$$
quite analogous to eq. \fsei.

Gravitino Pauli and quartic terms \susu\cremme\DFF\ are defined
by the formulas \LagV\ and \Lfourf\ with\foot{
The \K weights of the fermions are $p=-\bar p={1\over 2}$ for
$\psi_{\mu I}$, and  $p=-\bar p=-{1\over 2}$ for $\lambda^{iI}$.
The scalars and the fermions of the hypermultiplets, not discussed
here, have respectively \K weights $p=\bar p=0$ and
$p=-\bar p=-{1\over 2}$.}
$$ \eqalign{
Q^\Lambda&=L^\Lambda\ ;\qquad
P_\Lambda= M_\Lambda\cr
\cT^{\mu\nu}&=k_1 \bar\psi^I_\rho\psi^J_\s\e_{IJ}
\e^{\mu\nu\rho\s}
 }
\eqn\dulti
$$
for the purely gravitino terms, in which case the index $a$ of
the general treatment is obsolete.
For the mixed gaugino--gravitino Pauli terms we use
$$ \eqalign{
Q_\ib^\Lambda &=\cD_{\ib}\bar L^\L\ ;\qquad
P_{\Lambda\ib }= \cD_\ib\bar M_\L\cr
\cT^\ib_{\alpha\beta}&=k_2 \bar\l^\ib_I
\gamma_\rho\psi_{\s J}\e^{IJ}\e^{\mu\nu\rho\s}\ ,
 }
\eqn\nulti
$$
and the index $\ib$ plays again the role of $a$. The constants $k,
\, k_1$ and $k_2$ should also be fixed by supersymmetry. So, as
before, the unique quartic terms are generated by requiring duality
invariance of the action. Of course many of these terms are absent
in $N=1$ \cremme\ theories because of the absence of the second
gravitino. This is one of the differences between rigid
supersymmetry and local supersymmetry. What happens is that in $N=2$
supergravity, one introduces an extra $({3\over2},1)$ multiplet,
with respect to the $N=1$ case. This has the effect of having extra
auxiliary fields in the supergravity multiplet\ntSUGRA
$$
\cV^I_{J\mu}\ \ ,\ \ A_\mu\ \ ,\ \ T^-_{\mu\nu}\ \ , \ \ D
\eqn\cdinove
$$
other than the matter auxiliary field of the vector multiplet
$Y^{iIJ}$
(traceless, real, symmetric in $IJ$), $i,j=1,2$, \ie a real $SU(2)$
triplet.
 The
meaning of the auxiliary fields is straightforward. The $Y$'s
correspond to
the three auxiliary fields of a $N=1$ vector multiplet and a chiral
multiplet.
The $D$ auxiliary field gives the equation \csei\ (i.e  \csette ),
$T^-_{\mu\nu}$ is the graviphoton (symplectic invariant) combination
of the
gauge fields
$T^-_{\mu\nu}= T_\L\cF^{-\L}_{\mu\nu}$, and
 $\cV^I_{J\mu}\ , \ A_\mu $ are the composite $SU(2)$ and
$U(1)$ connections of the quaternionic manifold and K\"ahler--Hodge
manifold
 respectively.
Note that comparison between $N=1$ and $N=2$ theories shows that the
spinors
$\chi^i$ of the scalar multiplet and $\l^\Sigma$ of the vector
multiplet of
the $N=1$ theory are related to the doublet $\l^{iI}$ of the $N=2$
theory
by
$$
\chi^{i}=\l^{i1}\  \ \ ,\l^{\Sigma}=f_i^\Sigma \l^{i2}\ .
\eqn\cventi
$$

\section{ The three--form cohomology }

We recall that special geometry in
$N=2$ supergravity, unlike rigid special geometry, is suitable for
three--form cohomology for \cy manifolds. Let's define a holomorphic
three--form \doubref\specst\specco
$$
\Omega=X^\L\a_\L+F_\L\b^\L
\eqn\tuno
$$
 where $\a_\L,\b^\L$ is a $2n+2$
dimensional cohomology basis dual to the $2n+2$ homology cycles
($n=h_{21}$). $\Omega$ is a holomorphic section  of a line bundle.
Then it
follows that if one defines
$$
e^{-K}=i\int \Omega\wedge \bar\Omega >0
\eqn\tdue
$$
then
$$ g_{i\jb}={{-i\int \cD_i \Omega\wedge \cD_\jb \bar\Omega}\over i
\int
\Omega\wedge \bar \Omega}=-\del_i\del_\jb \log i \int \Omega\wedge
\bar\Omega
>0 \ .
\eqn\ttre
$$
 The $(2n+2)$ three--forms
$\cD_i\Omega,\cD_i\bar\Omega,\Omega,\bar\Omega$ with the cohomology
basis
$(\a_\L,\b^\L)$ correspond to the decomposition
$$
H^3(\IR)=H^{(2,1)}(\IC)+H^{(1,2)}(\IC)+H^{(3,0)}(\IC)+H^{(0,3)}(\IC)
\ .\eqn\tqua
$$
Note
that since $\Omega=(X^\L,F_\L)$, then $\cD_i\Omega=(\cD_iX^\L,\cD_i
F_\L)$, with
$f^\L_i=e^{{K\over2}}\cD_i X^\L,h_{i\L}=e^{{K\over2}}\cD_iF_\L$. The
relations
$$
 \int \Omega\wedge\Omega=\int\Omega\wedge\cD_i\bar\Omega=0
\eqn\tcin
$$
are obvious since
$\cD_i\Omega=\del_i\Omega-{{1}\over{(\Omega,\bar\Omega)}}
(\del_i\Omega,\bar\Omega)\Omega$. However the relation
$$
\int\Omega\wedge\cD_i\Omega=0 \ ,
\eqn\tsei
$$
 which is suitable for three--form
cohomology, implies
$$
 \int\Omega\wedge\del_i\Omega=0\ ,
\eqn\tset
$$
\ie
$$
 \del_i X^\L
F_\L-\del_i F_\L X^\L=0
\eqn\totto
$$
for any choice of the symplectic section. Eq. \totto\
is equivalent to
$$
X^\L\cD_i F_\L-\cD_i X^\L F_\L=0 \ .
\eqn\tnove
$$

\section{Duality transformations in $N=1$ locally supersymmetric
Yang--Mills theories}

In $N=1$ super Yang--Mills theories coupled to supergravity \cremme,
duality transformations are implemented as follows. Define the symplectic
$Sp(2 r)$ vectors
$$
\eqalign{
\cV &=(\cF_{\mu\nu}^{-A},G^{-\mu\nu}_A=
i{{\del \cL}\over{\del\cF_{\mu\nu}^{-A}}})\cr
\cU_\alpha &=(\l^A_\alpha ,f_{AB}(z) \l^B_\alpha )\cr}
\eqn\kuno
$$
where $(\l^A,\cF_{\mu\nu}^{-A})$ is the vector field strength multiplet
and $f_{AB}(z)$ is the holomorphic coupling  introduced in
\cremme\foot{We replaced the $f$ in \cremme\ by $2if$.},
which depends on the scalars of chiral multiplets, and which plays
here the role of $\bar {\cal N}_{AB}$ in the general treatment of
sections 2.1 and 2.2.
Then the $N=1$ supergravity lagrangian is  invariant under the
symplectic transformations
$$
\cV\to \cS \cV\ \ ,\ \ \cU\to \cS \cU\
 \ ,\ \ f\to(C+D f)(A+B f)^{-1}\ \ ,\ \ \cS\in
Sp(2,r;\IR)\ .
\eqn\ktre
$$
This is best seen using the $N=1$ tensor calculus (or superfield) notation
of ref. \cremme. The part of the action which contains the field strength
chiral multiplet
$$
W^A_\a=T(\cD_\a V^A)\ ,
\eqn\kquat
$$
where $T$ is the generalisation of to local supersymmetry of the
chiral projection $\bar D\bar D$ (similar to the operation obtaining
kinetic multiplets introduced in \pvan),
can be written in first order form by introducing an unconstrained chiral
multiplet $W^A_\a$ and a (vector) real lagrangian multiplier $U_A$
($f_{AB}$ is a chiral superfield)
$$
4\Im W^A_\a \cD_\b U_A \e^{\a\b}\mid_D+i f_{AB}(z) W_\a^A W_\b^B\e^{\a\b}
\mid_F\ .
\eqn\kcinque
$$
Variation with respect to $U_A$ yields the Bianchi identity
$$
\cD^\a W_\a^A=\bar \cD_{\dot \a} \bar W^{\dot\a A}\ ,
\eqn\ksei
$$
which is solved by
$$
W_\a^A=T(\cD_\a V^A)\ ,
\eqn\ksette
$$
which leads to the original form of the action.
The dual form of the theory is obtained,
 in a manner analogous to the rigid case \seiwit,
 by varying the same lagrangian with
respect to $W_\a^A$. Defining $W_{\a A}^{(D)}\equiv T(\cD_\a U_A)$,
and using the fact that the first term in  \kcinque\ can also be
written as $-2i  W_\a^A W_{\b B}^{(D)}\e^{\a\b} \mid_F$, yields
$$
W^A_\a=(f^{-1})^{AB} W_{\a B}^{(D)}\ ,
\eqn\kotto
$$
which implies the Bianchi identity also for $W^{(D)}$.
The dual lagrangian is
$$
\cL^D=-i (f^{-1})^{AB} W_{\a A}^{(D)}W_{\b B}^{(D)}\e^{\a\b}\mid_F\ .
\eqn\knove
$$
This realises the symplectic transformation of \ktre\ with
$B=-C=\bigone$ and $A=D=0$.

A duality rotation is a symmetry if for some coordinate changes $
z\to\tilde z$ ($z$ is the first component of a chiral multiplet)
$$
\tilde f_{AB}(\tilde z)=f_{AB}(\tilde z)
\eqn\kdieci
$$
and the superpotential W is a symplectic invariant section of a
Hodge bundle,
\ie
$$
\parallel \tilde W(\tilde z)\parallel^2=\parallel W(\tilde z)\parallel^2
 \ ,
\eqn\kundi
$$
where $\parallel W(z)\parallel^2=\mid W(z)\mid ^2 e^K\equiv e^G$.
In component form, we can exhibit the symplectic invariance of the
gaugino kinetic term and the Pauli terms by noticing that they
can be written as
$$
\eqalign{
e^{-1} \cL_{\rm kin}(\l,\bar\l) &=i \cU_\a\Omega{(\sigma^\mu)}^{\a\dot\a}
\cD_\mu\bar \cU_{\dot \a}\cr
e^{-1} \cL_{\rm Pauli} (\psi,\l) &=  \Im (\bar \cU_{\dot\a}\Omega
(\s^\mu)^{\dot\a\beta}\cV_{b\beta\gamma} \psi_\mu^\gamma )
 \cr
e^{-1} \cL_{\rm Pauli}(\chi,\l) &= \Im (\del_i f_{AB} \l^A_\a
\bar\chi^i_\b  \cF^{-B\a\b})\cr}
\eqn\kdodi
$$
where $\Omega$ is the symplectic metric $\pmatrix{0&1\cr -1&0\cr}$
(such that $\cS^T \Omega\cS=\Omega$)
and $\cV_b$ is the bare $\cV$ (only bosonic part).

The $(\psi,\lambda)$ Pauli term can be written in the form as in
\LagV\ and we identify in \elegant\  the symplectic vector $(Q,P)$
with $ \bar \cU_{\dot\a}$, and
$$
\cT ^{\dot \alpha}_{\beta\gamma}=-\ft12 (\s^\mu)^{\dot\a}{}_\beta
\psi_{\mu\gamma }\ .
$$
The last Pauli term,  $e^{-1} \cL_{\rm Pauli} (\chi,\l)$, has the
form \LagV, with
$$
H_{A\alpha\beta}=\ft12 \del_i f_{AB} \l^B_\a \bar\chi^i_\b \ .
$$
This we rewrite in the form \elegant using the following
identifications (note that $(\Im f)_{AB}$ is the matrix of the
kinetic terms of the vectors, and is thus invertible)
$$
\eqalign{
&Q^A_{i\alpha}\equiv (\Im f)^{-1\,AB} \partial_i f_{BC}
\lambda^C_\alpha \ ;\qquad
P_{Ai\alpha}\equiv \bar f_{AB} Q^B_{i\alpha} \cr
&\cT^{i\alpha}_{\beta\gamma}=\ft i4\delta^\alpha_{(\beta}
\chi_{\gamma)}^i\ . }
$$
To prove that these $(Q,P)$ form a symplectic vector, one uses
the following relations (which are in general true for
$f_{AB}$ replaced by $\bar {\cal N}_{AB}$):
$$
\eqalign{
\tilde f &=(C+Df)(A+Bf)^{-1}=(A^T+f B^T)^{-1}(C^T+f D^T)\cr
\del_i\tilde f &=D \partial_i f (A+B f)^{-1}-(C+Df)(A+Bf)^{-1}B
\del_i f(A+Bf)^{-1}\cr
&=(A^T+f B^T)^{-1} \del_i f(A+Bf)^{-1}    \cr
\Im \tilde f&=  (A^T+f B^T)^{-1}  (\Im f) (A+B\bar f)^{-1} \cr
\tilde\l &=(A+Bf)\l\cr }\ .
\eqn\ktredi
$$
These formulas then give automatically quartic fermionic terms as
discussed in section~2.

We observe that the requirements for having symplectic
transformations, \kdieci\ and \kundi, are in principle weaker than
what is necessary to have an $N=2$ theory.

\chap{Duality symmetries}
\section{The facts}
 Duality transformations in generic $N=2$ supergravity theories are a
different choice of the symplectic representative $(X^\L,F_\L)$ of
the
underlying special geometry. If the fields $\cF^{+\L}_{\mu\nu}
,G^{+}_{\L\mu\nu}$ have no electric
or magnetic sources these dualities are simply a different equivalent
choice of sections $(X^\L,F_\L)$ since they are defined up to a
symplectic transformation\susu\GAZU. However if the gauge fields are
coupled
to (abelian) sources then duality transformations map theories into
different theories with a duality transformed source. Since the
matrix
$\cN_{\L\S}$ plays the role of a coupling constant it is clear that
in
perturbation theory the only possible duality transformations are
those
with $B=0$ and have a lower triangular block form
$$
\cS=\pmatrix{A&0\cr C&A^{T-1}\cr}  \ .
\eqn\duno
$$
Under such change, the action changes in a total derivative which,
up to fermion terms, is
$$
\cL'(A,C)=\cL+\Im {\cal F}^{-\L}(C^T A)_{\L\S}{\cal F}^{-\S}  \ .
\eqn\ddue
$$
So the lagrangian is invariant up to a surface term. A duality
transformation is a symmetry if
$$
\tilde\cN(\tilde X,\tilde F)=\cN(\tilde X,\tilde F) \ .
\eqn\dtre
$$
If $F_\L=F_\L(X)$ this implies
$$
\tilde F(\tilde X)=F(\tilde X) \ .
\eqn\dquat
$$
Then using \dcin\ we should have \susu\villa
$$\eqalign{
2F[(A+B\cF )X] &=2F+2 X^\L (C^TB)_\L{}^\S F_\S\cr
{}~&+X^\L (C^TA)_{\L\S}X^\S+F_\L (D^TB)^{\L\S}F_\S\ ,\cr}
\eqn\dsei
$$
which is a functional relation for $F$ given $A,B,C,D$.
Note that because of \dcin\ it may happen that
$\tilde F (\tilde X)=0$. This is so when ${{\del
\tilde X^\L}\over{\del X^\S}}$
is not an invertible matrix.

\section{Heterotic $N=2$ superstring theories}

In  $N=2$ heterotic string theories, as the one obtained  by the
fermionic construction or by compactification on $T_2\times K_3$,
one often encounters
classical moduli spaces which are locally of the
form\nara\Shap\dive\abk\fepo\fklz
$$
{{O(2,n_v )}\over{O(2)\times O(n_v )}}
\times {{O(4,n_h)}\over{O(4)\times O(n_h)}} \ ,
\eqn\huno
$$
where $n_v$ and $n_h$ are respectively the number of the moduli in
vector
and hypermultiplets. If there are no charged massless
hypermultiplets
with respect to the gauge group $U(1)^r$, with $r=n_v$,
we may avoid holomorphic
anomalies \nrf{\DKL\CarOvr\fkdz\AGNT}\refs{\DKL{--}\AGNT}
 and the situation for this theory may be
similar to the rigid
Yang--Mills theory coupled to supergravity with an additional dilaton
axion multiplet.
According to the previous discussion,
all perturbative  duality symmetries are those for which the previous
formula
 holds
for a subgroup of lower triangular matrices
$$
\pmatrix{A&0\cr C& A^{T-1}\cr}
\eqn\hdue
$$
with $A^T C$ symmetric.

The $(r+2)\times (r+2)$ block $A$
contains the target space $T$ duality and $C$ contains the
Peccei--Quinn
axion symmetry \nfour\ (for the definition of $S$ in
the $N=2$ context, see below)
$$
S\to S+1\ .
\eqn\htre
$$
These are the tree level stringy symmetries of the massive states
with $M=|Z|$
 where $Z$ is the central charge of the $N=2$ supersymmetry algebra.
If the number of $T$--moduli is $r$ then the duality symmetries are in
$Sp(2r+4;\IZ)$.

 An important point is that we would like to make the tree level
 (string) symmetry manifest. This means that the gauge fields
$$
\cA^\L_\mu=(G_\mu,B_\mu,\cA^A_\mu)\ \ A=2,\ldots,r+1
\eqn\hqua
$$
($G_\mu$ is the graviphoton and the $B_\mu$ is the vector of the
dilaton--axion multiplet) should transform in the $2+r$ dimensional
(vector) representation of the target space duality symmetry
$$
\cA'=A \cA\ ;\qquad  A^T\eta A=\eta\ ;\qquad
\eta_{\L\Sigma}=Diag(1,1,-1,-1,\ldots)\ ,
\eqn\hcin
$$
with $A\in O(2, r;\IZ)$.
Under the axion Peccei--Quinn symmetry $S\to S+1$
$$
{\cA^\L} '=\cA^\L\ \ ,\ \  G_{\L\mu\nu}\to
G_{\L\mu\nu}+\eta_{\L\Sigma}{\cal F}^\Sigma
_{\mu\nu}\ ,
\eqn\hsei
$$
where
$$
\cN_{\L\Sigma}(S+1)=\cN_{\L\Sigma}(S)+\eta_{\L\Sigma}\ .
\eqn\hset
$$
This formulation is directly obtained by $N=2$ reduction of the
standard form of the $N=4$ supergravity action\nfour\fgkp\ with a
moduli space of the type \break
$O(6,r)/O(6)\times O(r)/\G$ and duality group $\G=O(6,r;\IZ)$.
However to get this in a standard $N=2$ supergravity form, one must
introduce $2+r$ symplectic sections $(X^\L,F_\L)$
($\L=0,1,\ldots,r+1)$ for
which $O(2,r)$ is block diagonal and the $S\to S+1$ shift is lower
triangular.
 This formulation can be obtained by making a symplectic
rotation, with $\cS$ given by
$$
\cS={1\over{\sqrt{2}}}\pmatrix{\bigone
&-\bigone\cr\bigone&\bigone\cr} \ ,
\eqn\hotto
$$
from a representation in which only $O(2)\times O(r)$ is block
diagonal \FRSO, namely
$$\eqalign{
O(2,r) :\ \ \
 \pmatrix{A&0\cr0&\eta A\eta\cr} &=\cS A_1 \cS^{-1}\cr
S\to S+1:\ \ \
 \pmatrix{\bigone &0\cr \eta &\bigone\cr} &=\cS A_2 \cS^{-1}\ ,
\cr}
\eqn\hnove
$$
where $A_1$, $A_2$ are the matrices given in ref. \FRSO.
The new sections
are given explicitly by eqs. \cquin,\csedi,
$$
\eqalign{
\hat X^\L &={1\over{\sqrt{2}}} (\d_{\L\S}-F_{\L\Sigma}) X^\Sigma\cr
\hat F_\L &={1\over{\sqrt{2}}} (\d_{\L\S} +F_{\L\Sigma}) X^\Sigma\ ,
\cr} \eqn\hundi
$$
where the function
$$
F=-\sqrt{X^2_i }\sqrt{X^2_\a}\ \ \ i=0,1;\ \ \ \a=2,\ldots,r+1
\eqn\hdodi
$$
was obtained in ref. \FRSO.
{}From \hundi,\hdodi\ one can verify that the  $\hat X^\L, \hat F_\L$
satisfy the constraints $\hat X^\L\eta_{\L\Sigma}\hat X^\Sigma=
\hat F_\L\eta^{\L\Sigma}\hat F_\Sigma= \hat X^\L\hat F_\L=0$.
In particular,
the new variables $\hat X^\L$ are not independent.
The previous constraints imply that we may set
$$\hat F_\L=S\eta_{\L\Sigma} \hat
X^\Sigma
\eqn\hhimpo
$$
and from eq. \dcin\ we find
$\hat F(\hat X)=0$. Note that this is precisely the case for which
$\hat F_\L=\hat F_\L(\hat X^\L)$ does not hold.

Since $O(2, r)$ is block diagonal,
 the new sections $(\hat X^\L,\hat F_\L)$ are $O(2, r)$
vectors. Recalling that the manifold ${{O(2,r)}\over
{O(2)\times O(r)}}$ can be described by the following
equations
$$\eqalign{
\eta_{\L\S} \Phi^\L \Phi^\S &=0\cr
\eta_{\L\S} \Phi^\L\bar\Phi^\S &=1\cr}
\eqn\hhdodi
$$
where $\Phi^\L$ are coordinates in $CP(1,r)$,
we may actually set
$$
\Phi^\L={{\hat X^\L}\over{\sqrt{\hat X^\Sigma \eta_{\Sigma\Pi}
\hat\bX{}^\Pi}}}\ .
\eqn\hquin
$$
 The \K potential is
$$
K=-\log i(\hat X^\L \hat \bF_\L-\hat \bX^\L\hat F_\L)=-\log i
(\bar S- S)- \log \hat X^\L  \eta_{\L\Sigma}\hat\bX^\Sigma \ .
\eqn\htredi
$$

Under $S\to S+1$
$$\eqalign{
\hat X^\L &\to \hat X^\L\cr
\hat F_\L &\to \hat F_\L+\eta_{\L\Sigma}\hat X^\Sigma\ .\cr}
\eqn\hdieci
$$
In the same basis the (non--perturbative) inversion $S\to -{1\over{S}}$
is given by the  symplectic matrix $\pmatrix{0& \eta\cr -\eta & 0\cr}$.
 This element, together with the one corresponding to $S\to S+1$
 generates an $Sl(2,\IZ)$ commuting with the $O(2r,\IZ)$ in $Sp(2r+4,\IZ)$.
The inversion is actually the only symmetry generator with $B\neq 0$.
It leaves invariant \htredi\ up to a \K transformation and it
 will be a symmetry of the classical spectrum (as it
comes by truncation of the $N=4$ spectrum \nfour) of electrically and
magnetically charged states discussed in chapter 5 .

The holomorphic sections $\hat X^\L$ can be written as  \fgkp
$$
\hat X^\L=({1\over2}(1+y^2_\a),{i\over2}(1-y^2_\a ),y^\a)\ ,
\eqn\hquat
$$
where the $y^\a$ are coordinates of the
$O(2, r)/O(2)\times O(r)$ manifold.
In terms of the $\Phi$ variables
the kinetic matrix $\hat\cN_{\L\Sigma}$ turns out to be
\fgkp\fepo\nfour
$$
\hat\cN_{\L\Sigma}(\hat X)=
(S-\bar S)
(\Phi_\L\bar\Phi_\Sigma +\bar\Phi_\L\Phi_\Sigma)+\bar S
\eta_{\L\Sigma} \ ,
\eqn\hsedi
$$
where $\Phi_\Lambda=\eta_{\Lambda\Sigma}\Phi^\Sigma$, and we will
also further raise or lower indices with $\eta$.

Notice that \hsedi\ cannot be computed directly from \ccin\ since
 in the new basis the
denominator identically vanishes. On the other hand, one can use
 the formula \bundi, which in our case becomes
$$
\hat\cN(\hat X,\hat F)=(\bigone+\cN (X))(\bigone-\cN(X))^{-1}
\eqn\hdicia
$$
and substitute for $X^\L$
the right hand side of the inverse transformations of \hundi
$$\eqalign{
X^\L &={1\over{\sqrt{2}}}(\delta_{\L\S}+S\eta_{\L\S})\hat X^\S\cr
F_\L &={1\over{\sqrt{2}}}(-\delta_{\L\S}+S\eta_{\L\S})\hat X^\S\ .\cr}
\eqn\hhtredi
$$

Formula \hsedi\ is precisely what is obtained from $N=4$ supergravity.
Because of target space duality we expect that also the $\hat
X^\L,\hat F_\L$
become, because of one loop corrections, a lower triangular
representation of $Sp(2r+4,\IZ)$
$$
\pmatrix{\hat X^\L \cr \hat F_\L\cr}\to
 \pmatrix{A&0\cr A^{T-1} C & A^{T-1}\cr}\pmatrix{ X^\L \cr F_\L}\ ,
\eqn\hdicio
$$
where the matrix $C$ comes from the monodromy of the one--loop term
\doubref\seiwit\KLTY.

It is interesting to compute explicitly the coupling of the dilaton to the
vector fields. The vector kinetic term is
$$
\Im \bar\cN_{\L\S}\cF^{-\L}_{\mu\nu} \cF^{-\S\mu\nu}=
-2\Im \cN_{\L\S}\cF^{\L}_{\mu\nu} \cF^{\S\mu\nu}+\Re \cN_{\L\S}
\cF^{\L\mu\nu}\tilde \cF^{\S}_{\mu\nu} \ .
\eqn\hhbas
$$
and, in particular, setting in \hquat\ $y^\a=0$, it becomes
$$
-2\Im S (\cF^0 \cF^0+\cF^1 \cF^1+\cF^\a \cF^\a)+
\Re S(\cF^0\tilde \cF^0 +\cF^1 \tilde \cF^1
-\cF^\a \tilde \cF^\a) \ .
\eqn\hhricc
$$
We see that the dilaton couples in a universal
 way to the vectors while in the topological term we have a coupling
with lorentzian signature.

\section{Duality symmetries in $N>2$ supergravities}

The general considerations of section 2 about duality symmetries
will apply to any higher $N>2$ extended supergravity theory.
Therefore, it is worth to briefly mention the implications of duality
symmetries for some non--perturbative properties that these theories
 may exhibit.
The important fact about $N>2$ theories is that the scalar field space is
(at least locally) a homogeneous symmetric space $G/H$, where $G$ is some
non compact subgroup of $Sp(2n)$ ($n$ is the total number of
vector fields existing in the theory). $H$ is its maximal compact
subgroup, as it must be for the kinetic matrix of the scalar field
space to be positive definite.

 On general grounds, we also know that the fields $(\cF^{-A},G^-_A)$
must belong to a linear representation of $G$ which is given
by the decomposition of the ($2n$--dimensional) vector representation
of $Sp(2n)$ under $G$. Thus, it is obvious that if this representation
remains irreducible in $G$, the duality symmetry will necessarily mix
electrically and magnetically charged states, since the $Sp(2n)$ vector
$(n^A_{(m)}=0,n^{(e)}_A)$ cannot be an invariant vector of $G$.

It is now
a fact of life that the full duality (continuous) symmetry $G$ of any
$N>2$ theory has a $2n$ dimensional representation which remains
irreducible under $Sp(2n)$  (see table below \salsez).
This immediately implies that, if we assume,
as conjectured in ref. \HuTo, that the full $G(\IZ)$ is a symmetry of the
dyonic states, then $G(\IZ)$ must be non--perturbative since the matrix
$B$ (see eq. \bcin ) in $G(\IZ)$ will not be vanishing. $N=3,5,6$
supergravities can be obtained as low energy limits of $d=4$ string
models \sgthree.

Another implication of this conjecture, for the case of $N=4$ theories,
is that, as pointed out in ref. \HuTo, the spectrum of the BPS states of the
ten dimensional heterotic string compactified on $T_6$ should be identical
to the spectrum of the same states for type II strings compactified on
$K_3\times T_2$, since the full $N=4$ BPS spectrum, invariant under
$Sl(2;\IZ)\times SO(6,n-6;\IZ)$ is completely fixed by supersymmetry.
This has the striking effect that at the non--perturbative level the type
II theory should exhibit enhanced gauge symmetries equivalent to the
$N=4$ heterotic string\foot{We acknowledge discussions with C. Hull on
this point.}.
\smallskip
\smallskip
\vbox{\tabskip=0pt \offinterlineskip
\def\tablerule{\noalign{\hrule}}
\halign to320pt{\strut#&\vrule#\tabskip=1em plus 2em&
\hfil#& \vrule#& \hfil#\hfil& \vrule#&
\hfil#& \vrule#\tabskip=0pt\cr\tablerule
&&N&& G                         && $repr.$ &\cr\tablerule
&&3&& $SU(3,n-3)$               && $(n_c)$ &\cr\tablerule
&&4&& $SU(1,1)\times SO(6,n-6)$ && $(2,n)$ &\cr\tablerule
&&5&& $SU(5,1)$                 && $(20)$  &\cr\tablerule
&&6&& $SO^*(12)$                && $(32)$  &\cr\tablerule
&&8&& $E_{7(7)}$                && $(56)$  &\cr\tablerule
}}
{Table: Representations of $G$ for $(F^{-\Lambda},
G^-_\Lambda)_{\Lambda=1, \ldots , n}$ in extended supergravities}

\chapter{On monodromies in string effective field theories}
\section{Classical and quantum monodromies}
We have just seen that the tree--level values of the
symplectic sections\break $(X^\L(z), F_\L(z))$ are given by
$$
X^\L\equiv X^\L_{\rm tree}\ \ \ , \ F_\L=S\eta_{\L\S} X^\S_{\rm tree}
\ .\eqn\muno
$$
The target space duality group $O(2,r;\IZ)$ acts non--trivially on
them
$$
\G_{\rm cl}:\ \ \ \pmatrix{X^\L \cr F_\L\cr}_{\rm tree} \to
\pmatrix{A &0\cr 0 & \eta A \eta \cr}\pmatrix{X^\L \cr F_\L\cr}_{\rm
tree}\ ,
\eqn\mdue
$$
generalizing the action of the Weyl group of the rigid case \KLTY.

At the one loop level, one expects that  $F_\L^{\rm tree}$ is changed
to \AGNT
$$
F_\L^{\rm tree} \to S X^\S \eta_{\L\S} +f_\L(X)
\eqn\mtre
$$
where $f_\L(X)$ is a modular covariant structure.

The associated perturbative monodromy can be obtained assuming,
according to
ref. \seiwit , that the rigid perturbative monodromy does not affect the
gravitational sector $X^0,X^1,F_0,F_1$. Thus the perturbative
lower triangular monodromy matrix is $\G_{\rm cl} T$, where\seiwit\KLTY
$$
T=\pmatrix{\bigone & 0\cr C & \bigone \cr}
\eqn\mquat
$$
and $C$ is
 an $(r+2)\times(r+2)$ symmetric matrix with non--vanishing
entries on the $r\times r$ block
$$
C=\pmatrix{\matrix{0&0\cr0&0\cr}&\matrix{\dots &0\cr\ldots &0\cr}\cr
\matrix{0&0\cr\vdots&\vdots\cr0&0\cr}& C_{ij}\cr}\ \ \ i,j=1,\ldots,r
\ .\eqn\mcin
$$
Indeed, we may think of decomposing $Sp(4+2r)$ into $Sp(4)\times
Sp(2r)$ and
simply assume that the rigid monodromy $\G_r \in Sp(2r)$ commute with
the gravitational $Sp(4)$ sector. This argument should at least apply
when the vectors of the Cartan subalgebra of the enhanced gauge
symmetry belong to the compact $O(r)$ in $O(2,r)$.

In string theory, the classical stringy moduli space
corresponds to the broken phase $U(1)^r$ of several gauge groups with
the
same rank. For instance, for $r=2$, $O(2, 2;\IZ)$ interpolates
between $SU(2)\times U(1)$, $SU(2)\times
SU(2)$ and $SU(3) $\veve. In the $N=4$ theory the $O(6;22)$ moduli
space
corresponds to broken phases of several gauge groups of rank $22$
such as,
$U(1)^6\times E_8\times E_8$ or $SO(32)\times U(1)^6$ or $SO(44)$
which are not subgroups one of the other \nara.

It is obvious that generically this means
that the one loop $\b$--function term \dive\seib\ should have
non--trivial monodromies at the
points where some higher symmetry is restored. For instance, for
$r=2$ we
may expect non trivial
 monodromies around $t=u$ $(SU(2)\times U(1)$ symmetry restored)
and $t=u=i$, $t=u=e^{2i\pi/3}$
 ($SU(2)\times SU(2)$ or $ SU(3)$ symmetry restored)
, $t,u$ being the parameters defined below.

This means that in supergravity theories derived from strings,
because of
target space T--duality, the enhanced symmetry points are richer than
in the
rigid case. Since different enhancement points are consequence of
$O(2,r;\IZ)$
duality, we expect that a modular invariant treatment of quantum
monodromies
will automatically ensure non trivial monodromy at the enhanced
symmetry
points.

In the sequel we shall discuss in some more detail the classical and
perturbative monodromies in the $r=1$ case ($O(2,1;\IZ)$) and the
classical
monodromies for $r=2$ ($O(2,2;\IZ)$).

Consider the tree level prepotential $F$ in the so--called cubic
form \susu\ for\break
${{SU(1,1)}\over{U(1)}}\times{{O(2,1)}\over{O(2)}}$ :
$$
F={1\over2} (X^0)^2 s t^2 \ ,
\eqn\msei
$$
where $s={{X^1}\over{X^0}}$ is the dilaton coordinate and
$t={{X^2}\over{X^0}}$
is the single modulus of the classical target space duality. We
parametrize
the $O(2,1;\IZ)$ vector as follows
$$
\eqalign{
X^0 &= {1\over2} (1-t^2) \cr
X^1 &= -t\ \ \ \ \ \ \ \ \ \ \ \  \ \ \ \ \ \ (X^0)^2+(X^1)^2-(X^2)^2=0\cr
X^2 &= -{1\over2}(1+t^2)\cr}
\eqn\mset
$$
The symplectic transformation relating $(X^\L,F_\L)$, ($\L=0,1,2$) to
the $(\hat X^\L,\hat F_\L)$ where $O(2,1)$ is linearly realized is
easily found to be
$$
\pmatrix{\hat X^\L \cr \hat F_\L\cr}=\pmatrix{P & -2R\cr R & P' \cr}
\pmatrix{X^\L \cr F_\L \cr}  \ ,
\eqn\motto
$$
where
$$
P=\pmatrix{{1\over2} & 0 & 0\cr0 &0& -1\cr -{1\over2} & 0 & 0\cr}\
;\qquad
P'=\pmatrix{1 & 0 & 0\cr0 &0& -1\cr -1 & 0 & 0\cr}\
;\qquad
R=\pmatrix{0 &{1\over2} & 0 \cr 0 & 0& 0\cr0 & {1\over2} &0}
\ . \eqn\mnove
$$
Let us now implement the $t$-modulus $Sl(2,\IZ)$ transformations
 $t\to-{1\over{t}}$, $t\to t+n$ (note that while $t\to -{1\over{t}}$
corresponds to the $SU(2)$ Weyl transformation of the rigid theory,
$t\to t+n$ has no counterpart in the rigid case, being of stringy
nature).
Using the parametrization \mset\ we find
$$
\eqalign{
t &\to -{1\over{t}} :\ \ \  \pmatrix{-1&0&0\cr 0&-1&0\cr
0&0&1\cr}\equiv -\eta\in O(2,1;\IZ)\cr
t &\to t+n :\ \ \ \pmatrix{1-{{n^2}\over2} & n & {{n^2}\over2} \cr
                           -n & 1 & n \cr
                            -{{n^2}\over2} & n & 1+{{n^2}\over2}
\cr}\equiv
V(n)\in O(2,1;\IZ)\ .\cr}
\eqn\mdieci
$$
Note that \mdieci\ implies $n\in 2\IZ$, i.e. the subgroup
$\G_{(0)}(2)$ of $SL(2,\IZ)$.
Actually this gives a projective representation in the subgroup in
$O(2,1;\IZ)$
of the matrices congruent to the identity $mod\ 2$.

It follows that $\G_{\rm cl}$ is generated by $(\G_1,\G_2)$ where
$$
\eqalign{
\G_1 &= \pmatrix{-\eta & 0 \cr 0 & -\eta \cr} \in Sp(6,\IZ)\cr
\G_2 &= \pmatrix{ V(2) & 0 \cr 0 &\eta V(2) \eta \cr}\in Sp(6,\IZ)
\ .\cr }
\eqn\mundi
$$
On the other hand it is possible to go to a stringy basis with a new
metric
$X^2_0 + X^2_1 - X^2_2 = \tilde X^2_1 +2  XY$
such that $SL(2,\IZ)$ is integral valued in $O(2,1;{\bf \IZ})$.

The $O(2,1;\IZ)$ generators corresponding to translation and
inversion are respectively given by:
$$
\pmatrix{1 &-2n &0\cr0 &1 &0\cr n & -n^2 & 1\cr}\ ;\qquad
\pmatrix{-1 & 0& 0 \cr 0 & 0 & -1\cr 0 & -1 & 0\cr}\ .
\eqn\orbi
$$

To make contact with the rigid theory it is convenient to define the
inversion
generator in $O(2,1;\IZ)$ with the opposite sign with respect to the
previous
 definition.

Let us now examine the perturbative monodromy matrices $T$.
 If we assume as
before that the $t\to -{1\over{t}}$ pertaining to the rigid theory
does not
affect the gravitational sector $(X^0,X^1,F_0,F_1)$, then we have
$$
T=\pmatrix{\eta & 0\cr C & \eta\cr} \ \ \ \ , \ \ C=\pmatrix{0&0&0\cr
0&0&0\cr 0&0&2\cr }
\eqn\mdodi
$$
corresponding to the embedding of the $Sp(2,\IZ)$ rigid
transformations
 acting on
the rigid section $(X^2,F_2)$ in $Sp(6,\IZ)$. Furthermore,
considering
the transformation of the $\cN_{\L\S}$ matrix and setting $D=A=\eta$
, $B=0$
we find
$$
\hat\cN_{22}=-2+\cN_{22}
\eqn\mtred
$$
for all other entries
$\hat {\cal N}_{\Lambda\Sigma}={\cal N}^{\Lambda\Sigma}$.
This is exactly the rigid result\seiwit.
However conjugating the $T$ matrix with $\G_2$
 one gets
$$
C_{\L\S}=\pmatrix{8&-8&-12\cr-8&8&12\cr-12&12&18\cr}
\eqn\mquattor
$$
which shows that $O(2,1;\IZ)$ introduces non--trivial perturbative
monodromies for all couplings. The other perturbative lower diagonal
monodromy is the dilaton shift \hnove\  which commutes with
$O(2,1;\IZ)$.

Analogous considerations hold for $O(2,n;\IZ)$, $n>1$. We limit
ourselves to write down the generators of $\G_{\rm cl}$ for the
$O(2,2;\IZ)$ case. We use the parametrization of $O(2,2)/O(2)\times
O(2)$ given by
$$
\eqalign{
X^0 &= {1\over2} (1-tu)\cr
X^1 &=-{1\over2} (t+u)\cr
X^2 &=-{1\over2} (1+ tu)\cr
X^3 &={1\over2} (t-u)\ \ \ \ \ \ \ \
(X^0)^2+(X^1)^2-(X^2)^2-(X^3)^2=0\ ,\cr}
\eqn\mquindi
$$
where $t,u$ are the moduli appearing in the $F$ function
$F= (X^0)^2\  stu$.
In the same way as for the $r=1$ case it is easy to find the
symplectic transformations relating the sections of the cubic
parametrization to the $X^\L$ defined in \mquindi.
They are given by
$$
\pmatrix{X\cr F\cr}\to
\pmatrix{ A& B\cr -B & A\cr}\pmatrix{X\cr F\cr}  \ ,
\eqn\trans
$$
with
$$\eqalign{
X &=(X^0, X^1, X^2, X^3)^T\ \ ,\ \ F=(F_0, F_1, F_2,  F_3 )^T\cr
 A &={1\over\sqrt{2}}\pmatrix{1&0&0&0\cr 0&0&-1&-1\cr -1&0&0&0 \cr
0&0&1&-1\cr}\ \
\ , \ \ B={1\over\sqrt{2}}\pmatrix{0&-1&0&0\cr 0&0&0&0\cr 0&-1&0&0\cr
0&0&0&0\cr}\ .\cr}
\eqn\matri
$$
 It is convenient to
use the string basis where the metric $\eta$ takes the form\nfour
$$
\eta=\pmatrix{0 & \bigone_{2\times 2}\cr \bigone_{2\times 2} & 0\cr}
\ ,\eqn\msedi
$$
corresponding to the basis ${1\over{\sqrt{2}}} (X^0\mp X^2),
{1\over{\sqrt{2}}} (X^1\mp X^3)$. Then one
finds the following $O(2,2;\IZ)$ representation
$$
\eqalign{
ut &\to +{1\over{ut}} : \pmatrix{0 & -\bigone \cr -\bigone &
0\cr}=\c_{ut}\cr
t &\to -{1\over{ t}} :\pmatrix{\epsilon & 0\cr 0 & \epsilon\cr}=\c_t \cr
u &\to -{1\over{u}} : \pmatrix{0 & \epsilon \cr \epsilon &
0\cr}=\c_u\cr
t &\to t+n : \pmatrix{ {\bf N}^t (-n) & 0\cr 0 & {\bf N} (n) \cr}
=\c_n\cr
t &\to u:  \pmatrix{a & b \cr b & a \cr} = \gamma\ ;\qquad
a = \pmatrix{1 & 0
\cr 0 & 0\cr}\ ;\qquad
b = \pmatrix{0 & 0 \cr 0 & 1 \cr}}
\eqn\mdicia
$$
where $\epsilon =\pmatrix{0&-1\cr 1&0\cr}$ and ${\bf N}(n)=
\pmatrix{1&n\cr 0&1\cr}$.

$\G_{\rm cl}$ is then generated by the matrices:
$$
\G_{ut}=\pmatrix{\c_{ut} & 0\cr 0 & \c_{ut}\cr} \ ;\
\G_t=\pmatrix{\c_t &0\cr 0 & \c_t\cr}\ ;\
\G_u=\pmatrix{\c_u & 0\cr 0 & \c_u\cr}\ ;
\G_n=\pmatrix{\c_n & 0\cr 0 & \c^T_{-n}\cr} \ .
\eqn\mdicio
$$
We note that the points $t=u$; $t=u=i$; $t=u=e^{{2 \pi i}\over{3}}$
are enhanced symmetry points corresponding to $SU(2)\times U(1)$,
 $SU(2)\times SU(2)$, and $SU(3)$ respectively \veve. Therefore we
 expect non--trivial quantum
 monodromies at these points according to the previous discussion.

\section{The BPS mass formula}

The classical and one loop monodromies are of course reflected in
symmetries of the electrically charged massive states belonging to
$O(2,n;\IZ)$ lorentzian lattice\nara. The BPS mass formula \Prso\
in the gravitational case is
$$
M=|Z|=|n^{(e)}_\L L^\L-n_{(m)}^\L M_\L|=e^{K/2}
 |n^{(e)}_\L X^\L-n_{(m)}^\L F_\L | \ .
\eqn\auno
$$
Note that the central charge $Z$ has definite $U(1)$ weight
$$
Z\to e^{(\bar f-f)/2} Z\ ,
\eqn\adue
$$
while the mass $M$ is \K invariant. The symplectic invariance of $M$
also
 implies that $(n^\L_{(m)},n^{(e)}_\L)$ transforms as $(X^\L,F_\L)$
$$
\pmatrix{n^\L_{(m)} \cr n^{(e)}_\L\cr}\to\pmatrix{A&B\cr C&D\cr}
\pmatrix{n^\L_{(m)} \cr n^{(e)}_\L\cr} \ ,
\eqn\atre
$$
where according to our previous discussion the perturbative
symmetries have $B=0$. Note that $n^\L_{(m)}, {n^{(e)}_\L}$ must
satisfy a lattice condition.
In the tree level approximation we may write
$$
M=|(n^{(e)}_\L-n^\S_{(m)} \eta_{\L\S} S) X^\L| e^{K/2}
\eqn\aqua
$$
which is invariant under the tree level symmetry $S\to S+1$, but also
under the non--perturbative inversion $S\to
-{1\over{S}}$\csf\dk\nfour\gh\ggpz\ taking into
account that
$$
K=- \log i (\bar S-S)-\log{{ X^\L \bar X^\S}\over{M_{Pl}^2}}
\eta_{\L\S}\ .
\eqn\acin
$$
Formula \aqua\ is therefore invariant under the $S-T$ duality
symmetry $Sl(2;\IZ)\times O(2,r; \IZ) \subset Sp(2r+4;\IZ)$.

The electric mass spectrum can be written as
$$
M^2_{(e)}
=|Z|^2= {{M^2_{Pl}}\over{2i (\bar S-S)}}  \cQ^{\L\S} n_\L^{(e)}
n_\S^{(e)}\ ,
\eqn\asei
$$
where $i(\bar S-S)={{8\pi}\over{g^2}}>0$ and
 $\cQ^{\L\S}=\Phi^\L\bar\Phi^\S+\bar\Phi^\L\Phi^\S$.
Formula \asei\ has exactly the same form as the analogous one
obtained in
$N=4$ (see ref \nfour).
When also magnetic charges are present, then
$$
\eqalign{
M^2 &= { {M^2_{Pl} }\over{i(\bar S-S)}}(n_\L^e-S n_\L^m)({1\over2}
\cQ_{\L\S}-{i\over2} \hat\cQ_{\L\S})(n_\S^e-\bar S n_\S^m)\cr ~
\ &={{M^2_{Pl}}\over 4}
(n_m,n_e) (\cM \cQ+\cL \hat\cQ)\pmatrix{n_m\cr n_e\cr}
\ ,\cr}
\eqn\asette
$$
where $\cM={1\over{\Im S}}\pmatrix{S\bar S&-\Re S\cr -\Re S & 1\cr}$,
$\cL=\pmatrix{0&-\bigone\cr \bigone & 0\cr}$ and
$\hat\cQ=i\left( \Phi^\L\bar\Phi^\S-\bar\Phi^\L\Phi^\S\right) $.
Recalling that
 $\cQ^{\L\S}={1\over2}(\eta^{\L\S}+{{\Im \cN^{\L\S}}\over{\Im S}})$,
this becomes
$$
\eqalign{
M^2 &= {{M^2_{Pl}}\over{i(\bar S-S)}}
 \ (n_\L^e-S n_\L^m)[{1\over4}
({ {\Im \cN_{\L\S}} \over {\Im S} }+\eta_{\L\S})
-{{i}\over2} \hat\cQ_{\L\S}](n_\S^e-\bar S n_\S^m)\cr
\  &={1\over4}
M^2_{Pl}(n_m,n_e) [{1\over2}\cM ({{\Im\cN}\over{\Im S}}
+\eta)+\cL\hat\cQ]\pmatrix{n_m\cr n_e\cr}\ .\cr}
\eqn\aatredi
$$
{}From this expression one can see that
the antisymmetric term $\hat\cQ$  vanishes if
$$
n^{(e)}_\L=m_1 n_\L\ \ ,\ \  n^\L_{(m)}=m_2 n_\S \eta^{\L\S}\ ,
\eqn\aaotto
$$
or, as it happens for the perturbative
 string, if no magnetic states are present
 ($n^m_\L=0\ ,
n^e_\L\equiv n_\L$).
In such case eq. \asette\ becomes
$$
M^2={{M^2_{Pl}}\over {8 \Im S}}|m_1-S m_2|^2[ n_\L n_\S(2\cQ^{\L\S}
-\eta^{\L\S})+n_\L n_\S \eta^{\L\S}]\ .
\eqn\aanove
$$
and  since $\Im \cN^{\L\S}$, being the vector kinetic matrix, is always
positive definite,
$$
M^2=0 \Longleftrightarrow n^\L n^\S\eta_{\L\S}<0\ \ (n^\L\neq 0)\ .
\eqn\aadieci
$$
As an example, take  $O(2,2;\IZ)$
and look for solutions of \aadieci\ corresponding to the string condition
$n^\L n_\L=-2$ . Using the parametrization
\mquindi\  we have
$$\eqalign{
n_\L X^\L &=n_0 X^0+n_1 X^1-n_2 X^2-n_3 X^3\cr
{}~ &={1\over 2}[(n_0+n_2)-(n_1+n_3)t-(n_1-n_3)u-(n_0-n_2)t u]\cr}
\eqn\aaundi
$$
Setting
$$
\eqalign{
n^0+n^2 &=-p_2 \sqrt{2}\cr
n^1+n^3 &=q_1 \sqrt{2}\cr
n^1-n^3 &= -p_1 \sqrt{2}\cr
n^0-n^2 &= q_2 \sqrt{2}\cr
n^\L n_\L &=(n^0+n^2)(n^0-n^2)+(n^1+n^3)(n^1-n^3)=-2(p_2 q_2+p_1 q_1)=-2\cr
{}~ & \to p_2 q_2+ p_1 q_1=1\cr}
\eqn\aaquattor
$$
we have
$$
n_\L X^\L={1\over{\sqrt{2}}}(-p_2-q_1 t+p_1 u-q_2 t u)\ .
\eqn\aasedi
$$
Let us verify that at the three enhancement points we get the correct
number of massless states.
If we take $t=u$ ($X^2=0$) we find
$$
\eqalign{
n_\L X^\L(t=u) &={1\over{\sqrt{2}}}[-p_2-(q_1-p_1)t-q_2 t^2]\cr
  &\to q_2=p_2=0\ \ \ q_1=p_1=\pm 1\ ,\cr}
$$
yielding the two massless states $(q_1,q_2)=(\pm 1,0)$.
In particular, for $t=u=i$ we have the solutions
$$\eqalign{
&n_\L X^\L(t=u=i)={1\over{\sqrt{2}}}[-p_2+q_2-(q_1-p_1)i]\cr
 &\to\ p_2=q_2\ \ ,\ \ q_1=p_1\ \ ,\ \ q_1^2+q_2^2=1\cr}
\eqn\aadicio
$$
yielding the four states $(q_1,q_2)=(\pm 1,0),(0,\pm 1)$. Taking instead
$t=u=e^{2\pi i/3}$ (such that $t^2=\bar t$),
we get
$$
\eqalign{
&n_\L X^\L(t=u=e^{2 \pi i/3})=0\cr
&\to +{1\over2}(q_1+q_2-p_1)-p_2=0\ \ ,\ \ q_1-q_2-p_1=0\cr
&\to p_1=q_1-q_2\ \ ,\ p_2=q_2\ \to\ q_1^2+q_2^2-q_1 q_2=1\cr}
\eqn\aadicia
$$
yielding the six states $(q_1,q_2)=(\pm 1,\pm 1),(\pm 1,0),(0,\pm 1)$.
As expected, these massless states together with the two original $(0,0)$
 states,
fill the adjoint representation
of $SU(2)\otimes U(1)$ ($t=u$), $SU(2)\otimes SU(2)$ ($t=u=i$),
$SU(3)$ ($t=u=e^{2 \pi i/3} $).

Unlike in $N=4$ theories, in $N=2$ theories the quantum spectrum
will not coincide with the classical spectrum. It will be found by
substituting $F_{\L {\rm tree}}\equiv S\eta_{\L\S}X^\S \to F_{\L
{\rm tree}}+{\rm quantum\; corrections}$ in \auno.

\chap{Conclusions}

In this paper we have formulated electromagnetic duality
transformations
in generic $ D=4$ , $ N=2 $ supergravities theories in a form suitable
to
investigate non--perturbative phaenomena.
Our formulation is manifestly duality covariant for the full
Lagrangian,
including fermionic terms, which unlike the rigid case, cannot be
retrieved
from the $ N=1$ formulation, nor from the  $ N=2 $ tensor calculus
approach.
Particular attention has been given to classical $T$-duality
symmetries
which actually occur in string compactifications and whose linear
action
on the gauge potential fields do not allow for the existence of a
prepotential
 function $F$ for the $N=2$ special geometry.
As examples we described the ``classical" electric and monopole
spectrum for
$T$--duality symmetries of the type $ O(2,r;\IZ)$, with particular
details
for the $r=1,2$ cases, by using the $N=2$ formalism.

For ``classical" monodromies this spectrum is of course related to the
spectrum
of $N=4$ theories studied by Sen and Schwarz \nfour.
Possible extensions of duality symmetries to type II strings have
been conjectured by Hull and Townsend \HuTo\ and also discussed in
\KLTY. In the present context
of $N=2$ heterotic strings the corresponding type II theories, having
$N=2$ space--time supersymmetry would correspond to $(2,2)$
superconformal field theories, i.e. quantum Calabi--Yau manifolds.

Due to the non--compact symmetries the BPS saturated states with
non--vanishing
central charges have a spectrum quite different from the rigid case.
Indeed
in rigid theories the ``classical" central charge $Z_{(cl)}$ vanishes
at the enhanced symmetry points where the original gauge group is
restored since
 there is no dimensional scale other than the Higgs v.e.v.. On the
contrary,
in the supergravity theory the BPS spectrum at these particular
points
corresponds in general to electrically and magnetically charged
states with Planckian mass (black holes, gravitational monopoles and
dyons)
\nrf{\gibhulman\asen\harliu\khu\kaor}
\refs{\gibhulman{,}\nfour{,}\asen{--}\kaor}. The only charged states
which become massless at the enhanced symmetric point are those with
$\eta^{\L\S} n_\L^{(e)} n_\S^{(e)}<0$.

We also discussed perturbative monodromies and their possible
relations
with the rigid case.
Non perturbative duality symmetries are more difficult to guess, but
it is
tempting to conjecture that a quantum monodromy consistent with
positivity
of the metric and special geometry may be originated by a
$3$-dimensional
Calabi-Yau manifold or its mirror image. If this is the case this
manifold
should embed in some sense the class of Riemann surfaces
studied\seiwit\KLTY\ in
connection
with the moduli space of $N=2$ rigid supersymmetric Yang-Mills
theories.

\noindent{\bf Acknowledgements.}

We would like to acknowledge useful discussions with the following
colleagues:
L. Alvarez-Gaum\'e, I. Antoniadis, M. Bianchi, M. Bill\`o,
 P. Fr\'e, C. Hull, E. Kiritsis,
C. Kounnas, W. Lerche, K. Narain, M. Porrati, T. Regge, G. Rossi
and T. Taylor.

\refout
\end